\documentclass[twocolumn]{aastex63}

\newcommand{\bprp}{$(G_{\rm BP} - G_{\rm RP})$}

\usepackage{isotope}

\received{\today}
\revised{---}
\accepted{---}
\submitjournal{ApJ}

\shorttitle{Modeling Gaia's M-dwarf Gap}
\shortauthors{Feiden, Skidmore, \& Jao}
\begin{document}

\title{Gaia Gaps and the Physics of Low-Mass Stars. I. The Fully Convective Boundary}

\correspondingauthor{Gregory Feiden}
\email{gregory.feiden@ung.edu}

\author[0000-0002-2012-7215]{Gregory A.\ Feiden}
\affiliation{Department of Physics \& Astronomy \\
             University of North Georgia \\
             Dahlonega, GA 30597, USA}
             
\author{Khian Skidmore}
\affiliation{Department of Physics \& Astronomy \\
             University of North Georgia \\ 
             Dahlonega, GA 30597, USA}
\affiliation{Department of Nuclear Engineering \\
             The University of Tennessee, Knoxville \\
             Knoxville, Tennessee 37996, USA}
             
\author[0000-0003-0193-2187]{Wei-Chun Jao}
\affiliation{Department of Physics \& Astronomy \\
             Georgia State University \\ 
             Atlanta, GA 30303, USA}

\nocollaboration{3}

\begin{abstract}
The Gaia M-dwarf gap is a significant under-density of stars observed near $M_G = 10.2$ in a color-magnitude diagram for stars within 200 pc of the Sun. It has been proposed that the gap is the manifestation of structural instabilities within stellar interiors due to non-equilibrium \isotope[3]{He} fusion prior to some stars becoming fully convective. To test this hypothesis, we use Dartmouth stellar evolution models, MARCS model atmospheres, and simple stellar population synthesis to create synthetic $M_G$-($G_{\rm BP} - G_{\rm RP})$ color-magnitude diagrams. We confirm that the proposed \isotope[3]{He} instability is responsible for the appearance of the M-dwarf gap. Our synthetic gap shows qualitatively similar features to the observed gap including: its vertical extent in $M_G$, its slope in the color-magnitude diagram, and its relative prominence at bluer colors as compared to redder colors. Furthermore, corresponding over-densities of stars above the gap are reproduced by the models. While qualitatively similar, the synthetic gap is approximately 0.2 magnitudes bluer and, accounting for this color offset, 0.16 magnitudes brighter than the observed gap. Our results reveal that the Gaia M dwarf gap is sensitive to conditions within cores of M dwarf stars, making the gap a powerful tool for testing the physics of M dwarf stars and potentially using M dwarfs to understand the local star formation history.
\end{abstract}


%%%%%%%%%%%%%%%%%%%%%%%%%%%%%%%%%%%%%%%%%%%%%%%%%%

\defcitealias{GS98}{GS98}
\defcitealias{GAS07}{GAS07}
\defcitealias{AGSS09}{AGSS09}

%%%%%%%%%%%%%%%%% BODY OF PAPER %%%%%%%%%%%%%%%%%%

\section{Introduction}
Launched in 2013, the Gaia space observatory's primary objective was to create the most comprehensive and precise catalog of stellar parallaxes and proper motions. The second major data release (DR2) provided precise parallax measurements to over 1 billion stars \citep{GaiaDR2}, increasing the number of stars with measured parallaxes by several orders of magnitude. The number of parallax measurements by Gaia allowed for the creation of more detailed color-magnitude diagrams \citep[CMD][]{GaiaHRD} than were possible with the several million stars in the Hipparcos catalogs \citep{Tycho2,Hipparcos2}. Among the new details revealed by Gaia was a narrow ``gap'' along the main sequence located among the mid-M dwarf stars \citep{Jao2018}.

The proposed gap in the Gaia DR2 CMD is a statistically significant drop in the density of stars compared to surrounding regions in the CMD \citep{Jao2018}. 
%A section of the Gaia DR2 CMD containing the gap is reproduced in Figure 1,  
The gap appears between $2.2 < (G_{BP} - G_{RP}) < 2.8$ with an absolute Gaia magnitude $10.0 < M_G < 10.3$. Closer analysis shows the density of stars drops by approximately $17\pm6$\% throughout the gap \citep{Jao2018}. In addition, a cross match between the Gaia DR2 \citep{GaiaDR2} and 2MASS catalogs \citep{2MASS} reveals that the gap is faintly present in the 2MASS data, but not at a significant level \citep{Jao2018}. This fact explains why the gap had previously gone undetected. Nevertheless, the faint 2MASS gap provides independent verification that the gap is not the result of a systematic error in the Gaia data.

The leading explanation for the gap is that it is a visible manifestation of the location where stellar interiors transition from partly to fully convective \citep{Jao2018, MacDonald2018, Baraffe2018}. Stars in the gap's vicinity have an M3V spectral type with an absolute $K_s$-band magnitude $M_{K} \approx 6.7$, which mass-luminosity relations suggest corresponds to stars with masses around $0.34 \pm 0.02\ M_{\odot}$ \citep{Benedict2016,Mann2019}. This is approximately the mass where stellar structure and evolution models predict the partly-to-fully convective transition occurs \citep{Chabrier1997, Baraffe1998}. Main sequence stars with masses above $\sim0.35\ M_{\odot}$ are predicted to be partly convective, while stars with masses below $\sim0.35\ M_{\odot}$ are predicted to be fully convective. The precise transition mass is model-dependent, but different models largely agree that the transition mass is $M = 0.35\pm0.05\ M_{\odot}$ \citep[e.g.,][]{Ezer1967, Copeland1970, Chabrier1997, Dotter2008, vanSaders2012, Chen2014}.

Regardless of the precise transition mass, stellar structure models predict that the transition is rather abrupt, occurring over a narrow mass range, $\Delta M < 0.05\ M_\odot$ \citep[see, e.g.,][]{Baraffe1998}. Typically, model grids are computed with a mass resolution of $\Delta M = 0.05$ or $0.10\ M_\odot$, meaning the transition occurs on a sub-grid scale, likely explaining why the transition was not apparent in theoretical luminosity functions. This characteristic agrees with what is observed in the Gaia DR2 gap; the gap occurs over a very narrow magnitude range, $\Delta M_K \sim 0.1$, which corresponds to a mass range $\Delta M \sim 0.02\ M_\odot$ \citep{Jao2018,Benedict2016,Mann2019}.

The physical explanation for why the transition from partly to fully convective interiors produces the observed Gaia gap is related to non-equilibrium \isotope[3]{He} fusion reactions as low-mass stars approach the zero-age main sequence \citep{MacDonald2018, Baraffe2018}. Non-equilibrium fusion of \isotope[3]{He} causes structural instabilities to occur over a narrow mass range near the fully convective boundary \citep{vanSaders2012}. While the fully convective boundary is often cited to occur around $M = 0.35\pm0.02\ M_{\odot}$, stellar structure and evolution calculations have long predicted that stars with masses $M \gtrsim 0.30\ M_{\odot}$ develop a small radiative shell separating a convective core and convective envelope as they arrive on the main sequence \citep[e.g.,][]{Ezer1967}. A convective core forms as the result of an overproduction of energy from the proton-proton (p-p) chain resulting from delayed ignition of \isotope[3]{He} $\rightarrow$ \isotope[4]{He} fusion relative to the first two fusion processes \citep{vanSaders2012, Baraffe2018}. With ignition of \isotope[3]{He} $\rightarrow$ \isotope[4]{He} fusion, the convective core grows in mass and radius until it merges with the convective envelope \citep{Ezer1967, vanSaders2012, Baraffe2018}. Rapid mixing of \isotope[3]{He} between the core and envelope causes the core \isotope[3]{He} abundance to decrease and fusion reactions to subside. The stellar core again collapses until \isotope[3]{He} is reestablished. This process occurs 4 -- 6 times before the star settles into fully convective equilibrium (see Section~\ref{sec:mechanism}) and leads to variability in stellar model radii and luminosities \citep{vanSaders2012}.

As a result of fluctuating stellar radii and luminosities, there is a decrease in the number density of stars at luminosities where the instability occurs \citep{MacDonald2018, Baraffe2018}. Stars within the \isotope[3]{He} instability region are spread over a larger area in the CMD compared to stars outside of the instability region. To date, this process has been confirmed to cause an under density in the theoretical luminosity function at solar metallicity \citep{MacDonald2018}. How the instability behaves in stellar evolution models as a function of metallicity remains unexplored. Furthermore, it has yet to be demonstrated that the \isotope[3]{He} instability in stellar models produces a characteristic gap across the entire main sequence. 

Accomplishing the latter task requires population synthesis to create a synthetic Gaia CMD from stellar evolution model predictions \citep{Baraffe2018}. If stellar evolution models are able to reproduce the observed gap morphology, it will provide confidence that the proposed physical mechanism --- van Saders's convective kissing instability \citep{vanSaders2012} --- is driving the observed Gaia M-dwarf gap. Such a result also acts as a test of the assumptions and approximations used for modeling nuclear reaction rates in low-mass stellar models.

In this study, we test whether stellar models can reproduce the Gaia M-dwarf gap's morphology in a synthetic CMD. The aim is to understand whether the \isotope[3]{He} instability's metallicity dependence is consistent with the gap's color-magnitude relationship. Consequently, we aim to test the hypothesis that the \isotope[3]{He} instability produces the observed gap. To do this, we use simple population synthesis to create a synthetic Gaia CMD that incorporates age and metallicity variations characteristic of the solar neighborhood. 
Our methods and assumptions are described in Section~\ref{sec:models}. A systematic analysis of our synthetic CMD and comparisons with the observed Gaia gap are presented in Section~\ref{sec:results}. In Section~\ref{sec:discussion}, we compare our results to previous studies, discuss model uncertainties, and discuss the potential to use the Gaia gap to constrain details about the Milky Way's star formation history.

\section{Modeling the Gap} \label{sec:models}
To generate a synthetic Gaia $M_G$/\bprp\ CMD, a simple population synthesis model was used to create a stellar population with theoretical properties determined from stellar structure models. Synthetic stellar properties were then transformed from the theoretical $L$/$T_{\rm eff}$ plane to a synthetic CMD using bolometric corrections from stellar model atmospheres. Details and assumptions for each step in the procedure are described below. Synthetic stellar populations and data generated for this study are available online.\footnote{\url{https://github.com/gfeiden/Gaia\_M-Dwarf\_Gap}}

\subsection{Simple Population Synthesis}
A population of synthetic stars characteristic of the solar neighborhood was created in multiple steps.
Masses between 0.1 $M_{\sun}$ and 0.8 $M_{\sun}$ were randomly drawn from a log-normal initial mass function \citep[IMF;][]{Chabrier2003}. The particular form of the IMF was not critical because the observed gap is a very narrow feature in the Gaia CMD and the appearance of the gap concerns the a relative measure of the local number density in the CMD. However, it is crucial that the mass distribution be continuous and smooth in the vicinity of $0.35 \pm 0.05 M_{\odot}$. A log-normal IMF with a characteristic mass of $0.08 M_\odot$ appears to provide an accurate description of the local field star population down to the hydrogen burning limit \citep{Chabrier2003}. 

Each mass was assigned a random metallicity drawn from a normal distribution with $[m/{\rm H}] = -0.14\pm0.19$~dex \citep{Nordstrom2004}. This distribution was a result from the Geneva-Copenhagen Survey, which measured spectroscopic metallicities for 14\,000 F and G stars in the local solar neighborhood. It is assumed that the local population of M dwarf stars, which is the focus of this study, largely follows the same abundance distribution as the local solar-like stars suitably corrected for diffusive processes. 

It is notable that surface abundances of F and G stars are expected to evolve in time as chemical elements undergo diffusion and settling processes \citep{Aller1960, Michaud1976, Thoul1994}. As a result, the metallicity distribution of local solar-like stars may be systematically biased toward lower metallicities. The magnitude of this effect is predicted from our models to be $|\Delta [\textrm{m/H}]| < 0.1$~dex for the ages and metallicities of stars under consideration. This effect is implicitly accounted for in the models, which are seeded with a higher initial $Z$ to ensure that a model of the Sun matches the adopted solar $Z/X$ at the solar age \citep[$\tau = 4.56$~Gyr; e.g.,][]{Connelly2012} with diffusion included \citep{Thoul1994}. Because low-mass stars have deep convective envelopes, they do not undergo any significant settling \citep[or no settling in the case of fully convective models;][]{Michaud1976}. Therefore, models of low-mass stars used in this study have a metallicity distribution of $[m/{\rm H}] = -0.09\pm0.19$~dex, consistent with trends in estimates of M dwarf metallicities \citep{RojasAyala2012, Mann2013, Birky2020}. 

Near the gap, a star's location in a CMD can be a sensitive function of age owing to the age sensitivity of the $^3$He instability \citep{vanSaders2012}. Given that the precise age distribution may affect the morphology of the gap, several distributions were adopted: (1) a uniform age distribution, (2) a normal distribution centered around the solar age, (3) a distribution skewed toward young, zero-age-main-sequence M dwarf stars, and (4) a distribution skewed toward older M dwarf stars. Results presented in Section~\ref{sec:results} adopt a uniform age distribution, but the impact of the precise age distribution is explored in Section~\ref{sec:discussion}.

Regardless of the precise distribution, synthetic stars were assigned ages between 500 Myr and 10 Gyr. There are two motivations for this range: (1) it corresponds to the range of ages where model stars undergo the $^3$He instability \citep{vanSaders2012, Baraffe2018}, and (2) it is roughly consistent with the ages of stars in the galactic thin disk \citep{Nordstrom2004, Holmberg2009, Kilic2017} and the maximum age of stars from the galactic thick disk \citep{Kilic2017}. While most of the stars in the local solar neighborhood likely belong to the thin disk, the long lifetimes of M dwarfs suggest that there may be a non-negligible fraction of thick disk interlopers \citep{Laughlin1997}. 

\subsection{Stellar Evolution Models}
Stellar properties were determined from the random combinations of mass, metallicity, and age by interpolating within a fine grid of stellar evolution model isochrones. Adopted models are based on a modified version of the Dartmouth Stellar Evolution Program \citep{Dotter2008, Feiden2016}. A complete description of the input physics relevant for low-mass stars and of all modifications to the original Dartmouth model physics will be presented in a separate model release paper (Feiden et al., in prep.).

Two grids of standard stellar evolution model isochrones were utilized. One grid adopted the solar composition of \citet[hereafter GS98]{GS98} while the other adopted the composition of \citet[hereafter GAS07]{GAS07}. The \citetalias{GS98} solar composition has a higher overall metal mass fraction, $Z = 0.018$, than \citetalias{GAS07} ($Z = 0.013$), leading to larger radiative opacities in stellar structure models. As a consequence, standard solar models using a \citetalias{GS98} composition are better able to predict the helioseismic sound speed profile of the Sun, especially near the base of the solar convection zone, than if a \citetalias{GAS07} composition is adopted \citep{Bahcall2005, Basu2008}. For this reason, \citetalias{GS98} abundances can be viewed as more favorable by the stellar interior community. However, modern measurements and analysis techniques continue to support an atmospheric solar composition that is similar to --- albeit slightly more metal rich than --- the \citetalias{GAS07} composition \citep{AGSS09, Caffau2011}. While two grids were created, results presented in Section~\ref{sec:results} were generated with the \citetalias{GAS07} grid. A comparison is presented in Section~\ref{sec:discussion}, but there are essentially no significant differences with regard to the Gaia M dwarf gap.

The choice of a \citetalias{GAS07} solar composition over more recent determinations \citet[hereafter AGSS09]{AGSS09} is motivated by the fact that MARCS model atmospheres \citep{Gustafsson2008}, used to specify surface boundary conditions, adopt the \citetalias{GAS07} solar composition. Model atmosphere structures and synthetic spectra are not significantly impacted by the choice of a \citetalias{GAS07} or an \citetalias{AGSS09} composition (B.\ Edvardsson, private communication). Comparisons between stellar evolution models computed with MARCS/\citetalias{GAS07} and PHOENIX/\citetalias{AGSS09} surface boundary conditions show negligible structural differences.

Each grid was computed with mass tracks between 0.08~$M_\sun$ and 1.0~$M_\sun$. A mass increment of 0.005~$M_\sun$ was used for 0.08 -- 0.20~$M_\sun$, 0.01~$M_\sun$ for 0.20 -- 0.28~$M_\sun$, 0.005~$M_\sun$ for 0.28 -- 0.43 $M_\sun$, and 0.02~$M_\sun$ above 0.36~$M_\sun$. The two highest resolution domains were chosen to given adequate resolution in effective temperature below 0.2~$M_\sun$ and to adequately resolve the $^3$He instability at various metallicities near 0.34 $M_\sun$ \citep{vanSaders2012, MacDonald2018}. Mass tracks were computed with metallicities $-0.7$~dex $\le$ [$m$/H] $\le$ $+0.5$~dex with 0.1~dex resolution.

Stellar model isochrones were created for ages between 500 Myr and 10 Gyr with an age resolution of 25 Myr between 500 Myr and 1 Gyr, and 100 Myr resolution between 1 Gyr and 10 Gyr. Model properties for stellar masses at each age were extracted from the mass tracks using linear interpolation. Because K and M dwarf stars evolve slowly, there are no complex evolutionary features to capture that require sophisticated isochrone generation methods. Isochrones tabulate model masses, effective temperatures, surface gravities, bolometric luminosities, radii, lithium abundances, convective envelope masses, surface magnetic field strengths (for magnetic models), and the convective overturn times.

With a grid of stellar model isochrones, a 3-dimensional linear interpolation is used to assign predicted model properties to the random combination of masses, metallicities, and ages compiled by the simple stellar population model.  

\subsection{Synthetic Photometry}
Synthetic photometric magnitudes were calculated using synthetic spectra produced from MARCS model atmospheres \citep{Gustafsson2008} using Gaia zero-points established by \citep{Casagrande2014, Casagrande2018}. Because MARCS specializes in the computation of cool star atmospheres, MARCS does not have a model of Vega with which to derive zero-points. For this reason, special care needs to be taken to ensure reliable MARCS synthetic photometry \citep{Casagrande2014}. There are multiple transmission profiles available for the Gaia photometric passbands. We used the revised Gaia DR2 passband profiles \citep{GaiaPhot}.\footnote{https://www.cosmos.esa.int/web/gaia/iow\_20180316} Systematic errors in the derived passbands and variations between different estimates for the photometric passbands can reach up to 10 millimagnitudes \citep{GaiaPhot}, but this does not significantly affect the key results of our work. 

\begin{figure*}[ht]
    \centering
    \includegraphics[width=1.70\columnwidth]{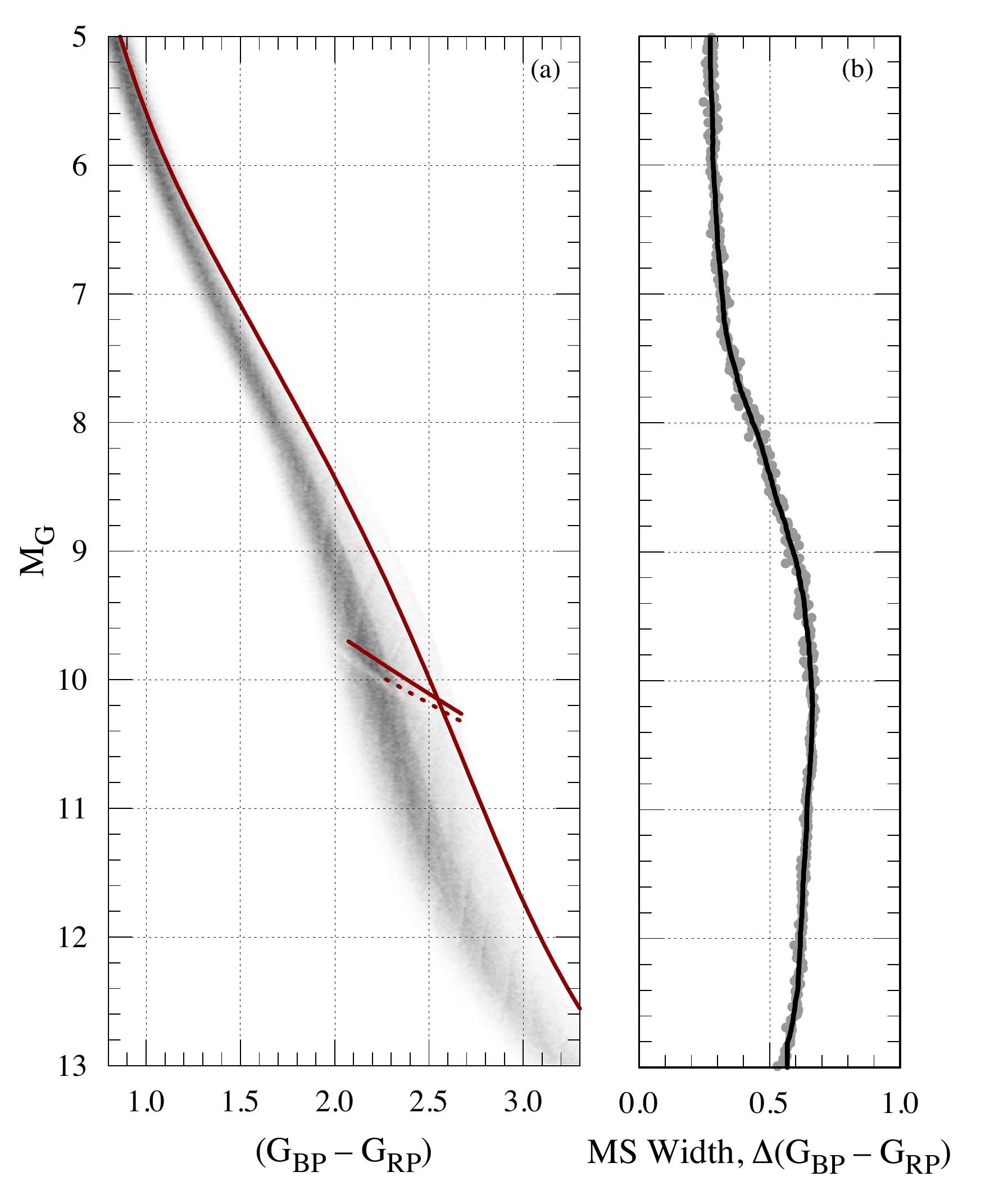}
    \caption{(a) Synthetic Gaia color-magnitude diagram created using simple population synthesis. Densities of stars are shown as a 2D histogram using a grey-scale. The bin resolution is 0.02 magnitudes in $M_G$ and 0.02 magnitudes in \bprp. Red solid lines indicate the observed and dereddened main-sequence locus and bright edge of the M dwarf gap \citep{Jao2020}, while the red dotted line indicates the observed gap's $M_G$ of greatest decrement \citep{Jao2018}. (b) The synthetic main-sequence \bprp\ width as a function of $M_G$. Gray points are width measurements every 0.01 magnitudes in $M_G$. The solid line is a moving average with a boxcar kernel with a width of 0.2 magnitudes.}
    \label{fig:gaia_cmd}
\end{figure*}

\section{Results} \label{sec:results}

\subsection{Color-Magnitude Diagram}
A synthetic Gaia $M_G$/\bprp\ CMD is shown in Figure~\ref{fig:gaia_cmd}(a). The CMD is divided into bins of 0.02 magnitude in color and absolute magnitude with the gray shading indicating the number of stars in each bin. A darker shade implies more stars in a given bin. The single star main sequence has a prominent spine as a function of $M_G$ with stars distributed in color around the spine (recall that we ignored binary systems). There is some width to the main sequence owing to metallicity variations and, to a lesser extent, age variations in the synthetic stellar population. The prominent spine to the main sequence corresponds to the peak of the metallicity distribution and disappears if a uniform metallicity distribution is adopted (see Section~\ref{sec:feh}). The main sequence appears to extend over a larger color domain on the red side of the main sequence spine than on the blue, a result of the dependence of M dwarf star colors on metallicity \citep{Allard1995}. Similar features are observed in the Gaia CMD \citep{Jao2018, Jao2020}, suggesting that the choice of a non-uniform metallicity distribution was appropriate. 

While the synthetic CMD is qualitatively similar to the observed CMD, it's clear that the two CMD morphologies are not quantitatively similar. This is illustrated in Figure~\ref{fig:gaia_cmd}(a) by comparing the synthetic CMD to the red solid and dashed lines. These lines represent the observed Gaia CMD main sequence spine \citep{Jao2020}, the bright edge of gap \citep{Jao2020}, and the magnitudes of greatest decrement within the gap \citep{Jao2018}. Most obvious is that the synthetic CMD is bluer than the observed CMD by approximately 0.3 magnitudes in \bprp\ at a given absolute magnitude. The synthetic CMD also appears to be brighter than the observed CMD by about 0.15 magnitudes in $M_G$ based on the location where the M dwarf gap crosses the main sequence spine (Section~\ref{sec:gap}). The offset is color-dependent, disappearing for K dwarfs with absolute magnitudes $M_G < 6$. Remaining offsets in the K-dwarf regime can potentially be attributed to photometric zero-point errors, Gaia transmission profile uncertainties, and reddening errors. Disagreement at fainter absolute magnitudes is not surprising. It is well documented that stellar structure models adopting synthetic color transformations consistently predict M star optical colors that are too blue compared to observed CMDs 
\citep[e.g.,][]{Naylor2009, Bell2012, Rajpurohit2013}. Despite the inability of models to predict accurate colors and magnitudes of M dwarf stars, models do better when predicting relative colors of stars when accounting for different metallicities, which manifests itself in the CMD's width.

The synthetic CMD's main sequence width as a function of $M_G$ is shown in Figure~\ref{fig:gaia_cmd}(b).
The K dwarf main sequence has a consistent width of 0.27 magnitudes down to $M_G = 7.2$. At absolute magnitudes fainter than $M_G = 7.2$ (late K), the main sequence starts to broaden in \bprp. It reaches a maximum width in \bprp\ of 0.66 magnitudes between $M_G = 10.05$ and $M_G = 10.35$, beyond which the main sequence begins to narrow. The magnitude range over which the main sequence broadens corresponds to where TiO molecular bands are strengthening and having an impact on M dwarf optical colors \citep{Boeshaar1976, Allard1995}. Because the abundance of TiO is strongly dependent on metallicity, M dwarf optical colors become increasingly sensitive to metallicity, leading to the observed main sequence broadening \citep{Allard1995}. At magnitudes fainter than $M_G = 10.4$, increased H$_2$O opacity in the near-infrared and a saturation of TiO bands in the optical diminish the role of metallicity in dictating observed M dwarf optical colors \citep{Allard1994}. While CaOH and VO are beginning to emerge in M dwarf spectra at this point, their role in dictating the metallicity-dependence of stellar colors is not as strong as the role of TiO \citep{Allard1995}.

The single-star main sequence width is dependent on the metallicity range adopted in the population synthesis procedure and the metallicity distribution of stars in the solar neighborhood. Therefore, one cannot necessarily expect exact quantitative agreement between the synthetic main sequence and the observed Gaia main sequence. However, the two CMDs show qualitative agreement in that both start narrow for brighter absolute magnitudes in the K dwarf regime and then progressively broaden toward fainter magnitudes. The observed Gaia CMD has an apparent maximum width around $M_G \sim 10.5$ \citep{Jao2020}. It is remarkable, however, that the maximum width of the observed Gaia CMD is about 0.7 magnitudes in \bprp\ \citep{Jao2018,Jao2020}, which compares well with the synthetic value of $0.66$ magnitudes in \bprp. Agreement between main sequence widths may hint at a lack of metallicity dependence in observed color offsets between the observed and synthetic CMD. If confirmed, this could suggest that continuous opacities are primarily responsible for the observed offset. 

\subsection{M-dwarf gap morphology}
\label{sec:gap}

\begin{figure}
    \centering
    \includegraphics[width=\columnwidth]{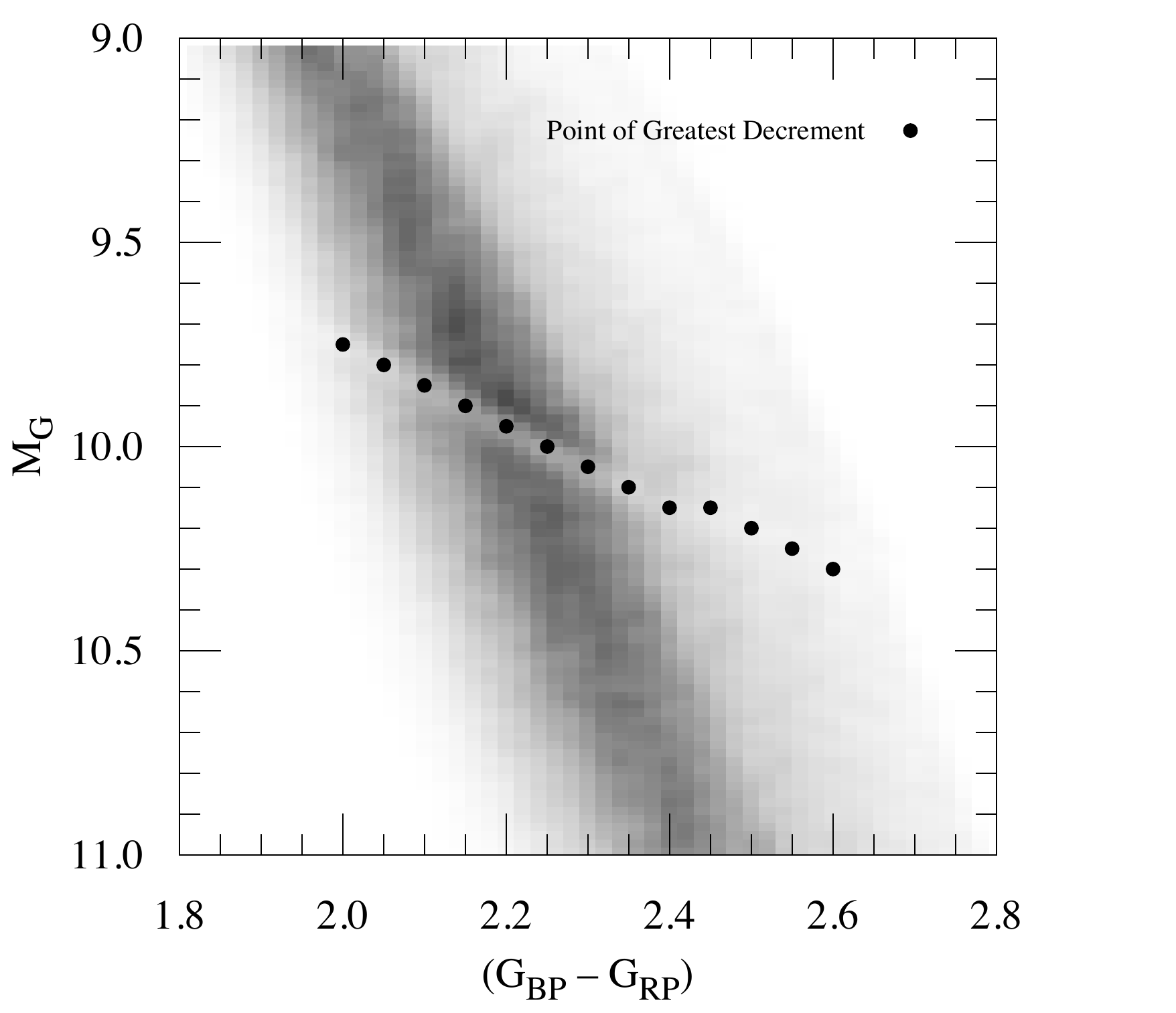}
    \caption{View of the region surrounding the Gaia M-dwarf gap in a synthetic color-magnitude diagram (CMD). Densities of stars are shown as a 2D histogram using a grey-scale with bins of $0.02 \times 0.02$ magnitudes. Filled black circles represent locations in the gap that show the greatest under density of stars compared to surrounding regions in the CMD.}
    \label{fig:gap_slopes}
\end{figure}

\begin{figure}
    \centering
    \includegraphics[width=\columnwidth]{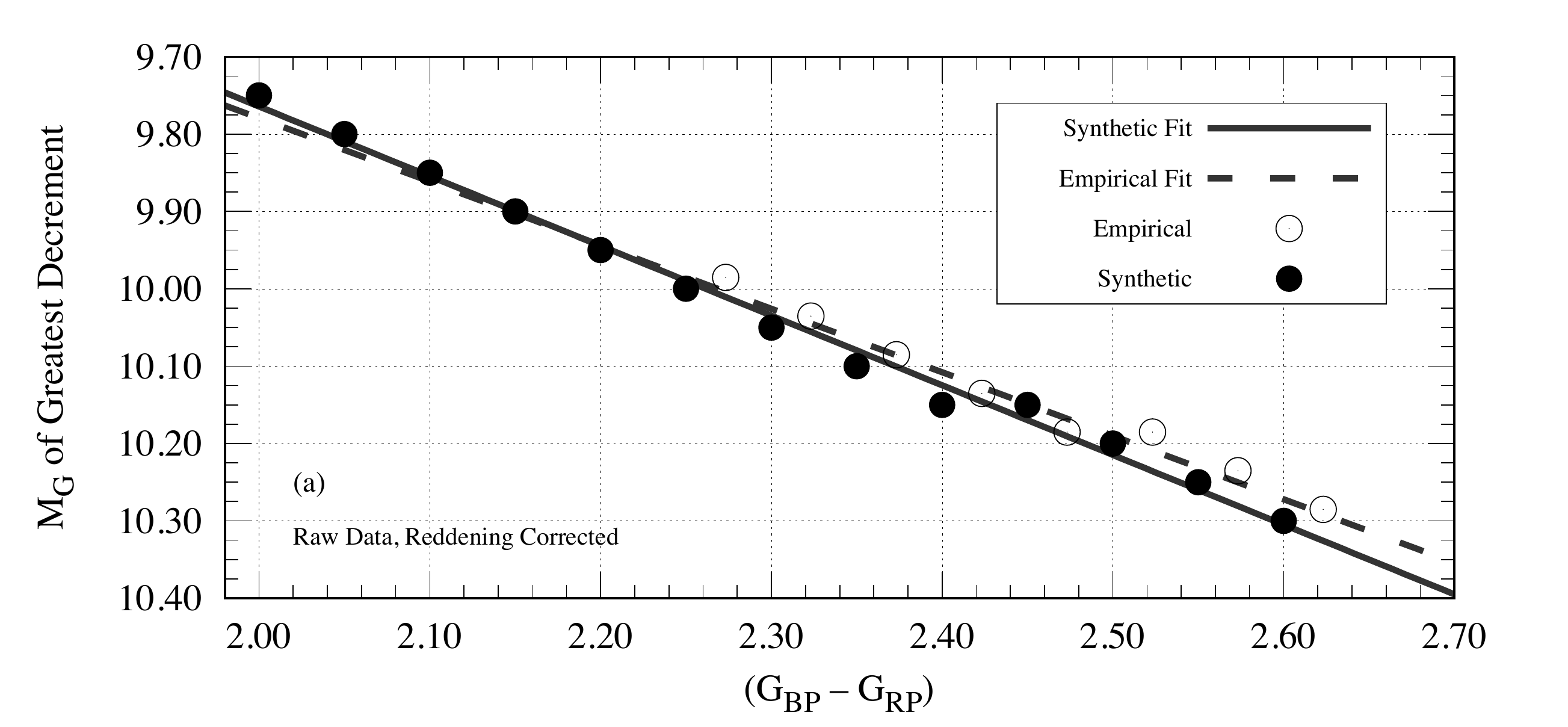} \\
    \includegraphics[width=\columnwidth]{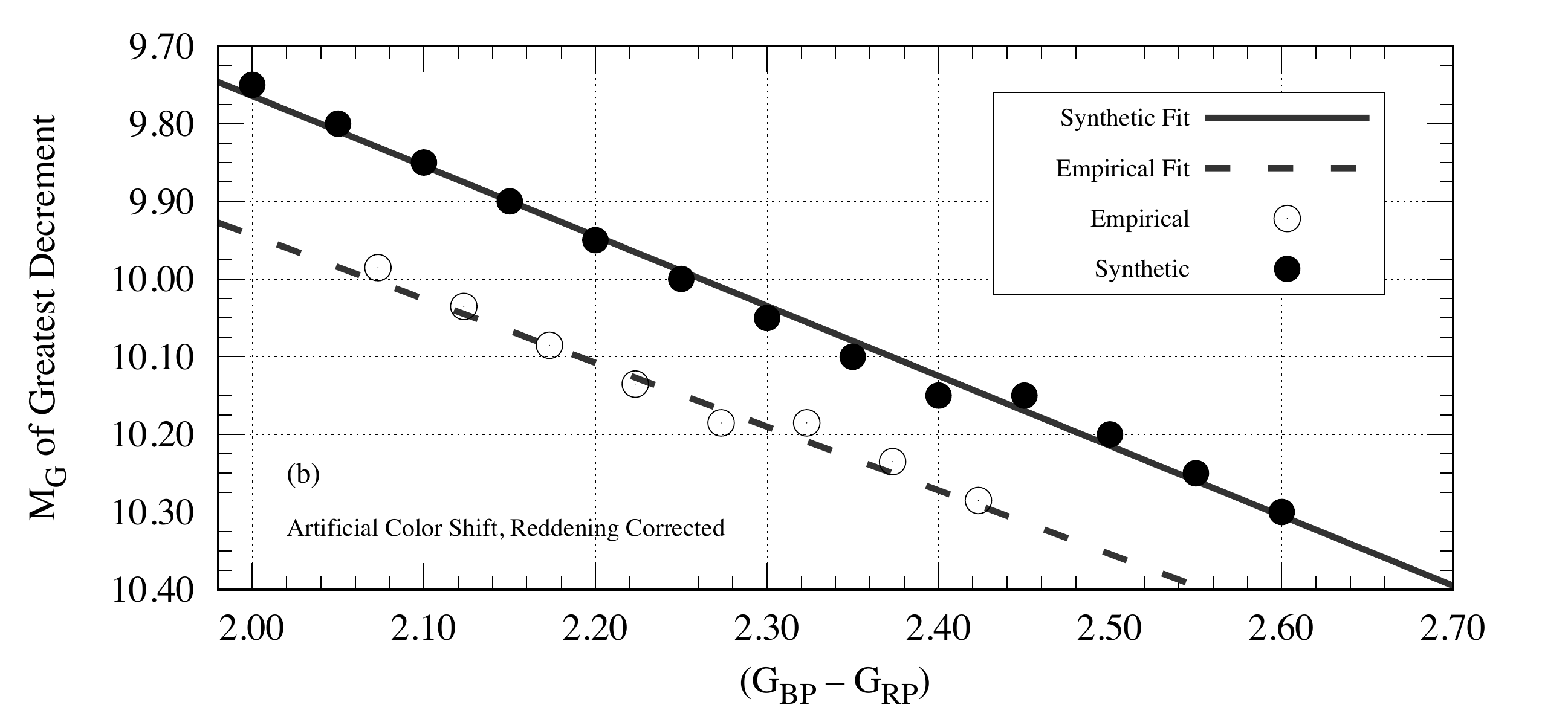}
    \caption{(a) Absolute magnitude, $M_G$, where the distribution of stars in a given color bin exhibit the greatest difference between the observed number and that expected from a skewed Gaussian fit to the total distribution of stars. Empirical data \citep[open circles;][]{Jao2018} and synthetic results (filled circles) are shown along with linear fits to the data (dashed and solid lines, respectively). (b) Same as the top, but with the empirical data shifted by 0.20 magnitudes in \bprp, which corresponds to the color difference between synthetic and empirical color-temperature transformations (see Section~\ref{sec:gap}).}
    \label{fig:regression}
\end{figure}

To identify and characterize the M-dwarf gap, the synthetic CMD was divided into square bins of 0.05 magnitudes. For each color bin between \bprp\ = 1.95 -- 2.60, a skewed Gaussian profile was fit to the distribution of stars in $M_G$. The gap's location in each color bin was defined to be the point where the residuals between the best-fit skewed Gaussian and the actual $M_G$ distribution was greatest.\footnote{This is similar to the procedure adopted by \citet[]{Jao2018}, although they fit a standard Gaussian distribution.} 
These points are presented in Figures~\ref{fig:gap_slopes} and \ref{fig:regression}, and are hereafter referred to as points of greatest decrement.

The synthetic gap has an average $\langle M_G\rangle = 10.0$. This compares favorably to the observed gap, which has an average $\langle M_G\rangle = 10.1$ \citep{Jao2020}. Despite the average (and median) values being within 0.1 magnitude, a more accurate estimate of the absolute magnitude difference is 0.15 magnitudes after correcting for the noted color offset (see Figure~\ref{fig:regression}). This represents a 2 - 3\% difference between the gap's predicted and observed absolute magnitude in $M_G$. A similar offset was observed between stellar evolution model predictions and the observed mass-$M_K$ relationship \citep{Mann2019}. Whether these offsets represent an offset in stellar model predictions of M dwarf bolometric luminosities or a consistent offset in color-temperature transformation remains unclear.

Despite reasonable agreement between the observed and predicted gap location in $M_G$, the synthetic main sequence --- and thus the gap's location in \bprp\ --- is offset toward bluer colors. There are multiple ways to estimate this offset. For example, the observed gap intersects main sequence locus at (\bprp, $M_G$) = (2.60, 10.3), whereas the synthetic gap crosses the synthetic main sequence at (\bprp, $M_G$) = (2.22, 9.90). This suggests the synthetic gap is 0.38 magnitudes bluer and 0.40 magnitudes brighter than expected.
Furthermore, the greatest under density of stars in the observed Gaia CMD occurs around \bprp\ $\sim 2.4$. However, models predict the greatest under density at \bprp\ $\sim 2.05$, which is 0.35 magnitudes bluer than the observed \bprp\ color, consistent with the offset estimated from the gap's intersection with the main sequence. Both the synthetic and observed CMDs show that the largest under density of stars is approximately 30\%. 

Alternatively, we can compar synthetic colors against empirical estimates. Doing so suggests that our predictions are too blue by about 0.2 magnitudes \citep{Pecaut2013}.\footnote{see \url{http://www.pas.rochester.edu/~emamajek/EEM_dwarf_UBVIJHK_colors_Teff.txt}} Shifting the observed CMD 0.2 magnitudes bluer in \bprp\ leads to the observation that the observed and synthetic gaps are offset by 0.16 magnitudes in $M_G$ (as noted above). This is illustrated in Figure~\ref{fig:regression}, where the magnitudes of greatest decrement are plotted against \bprp\ color. We favor this estimate of the synthetic-observed gap offset because changes to synthetic model atmospheres or stellar interiors that can mitigate observed color offsets are likely to impact characteristics of the gap, such as those used in our analysis above. Curiously, this exercise highlights that the observations support a model prediction that the magnitude of greatest decrement is directly proportional to the \bprp\ color, but plateaus in the middle of the gap. 

Across the synthetic CMD, the gap's height is approximately 0.05 -- 0.15 magnitudes in $M_G$. The gap appears to extend over 0.15 magnitudes in $M_G$ at colors bluer than \bprp\ $\sim 2.25$, and narrows to 0.05 magnitudes in $M_G$ at the reddest colors. Originally, the gap was observed to have a very narrow height of 0.05 magnitudes in $M_G$, but revised estimates indicate that the gap extends over a greater $M_G$ range at bluer colors as compared to redder colors \citep{Jao2020}, consistent with model predictions. %% CHECK THIS

Models predict that the gap has a slope of $0.90\pm0.02$ in the synthetic Gaia CMD. The slope was obtained by performing a linear regression on the points of greatest decrement presented in Figure~\ref{fig:gap_slopes}. The fit result is shown in Figure~\ref{fig:regression}. The points of greatest decrement in the synthetic CMD appear to trace the gap's faint edge. By comparison, the slope of the observed gap is $0.82\pm0.05$ \citep{Jao2018}. Using the updated gap location from \citet[]{Jao2020}, the gap's slope is estimated to be $0.941 \pm 0.002$. Both values must be interpreted with caution. The gap from \citet[]{Jao2018} only spans a color domain of 0.40 magnitudes, compared to the synthetic gap (0.65 magnitudes) and the enhanced Gaia CMD \citep[0.60 magnitudes;][]{Jao2020}. A narrower color domain places additional weight on the plateau feature observed in Figure~\ref{fig:regression}. Points from the revised gap location span a similar color domain \citep{Jao2020}, but trace the gap's bright edge instead of the location of greatest decrement. Nevertheless, the fact that the synthetic gap slope lies between these values and reproduces the slope on either side of the plateau in Figure~\ref{fig:regression}, suggests consistency between the model and observed gap. % interior physics!!!

\subsection{Gap Formation}

\subsubsection{Origin of the Gap}
A proposed explanation for the gap is that stars in and around the gap are undergoing a transition from having partly to fully convective interiors. The gap is generated by a structural instability caused by non-equilibrium \isotope[3]{He} fusion \citep{vanSaders2012, MacDonald2018, Baraffe2018}. The synthetic stellar population shown in Figure~\ref{fig:gaia_cmd} supports this hypothesis.

\begin{figure}[t]
    \centering
    \includegraphics[width=\columnwidth]{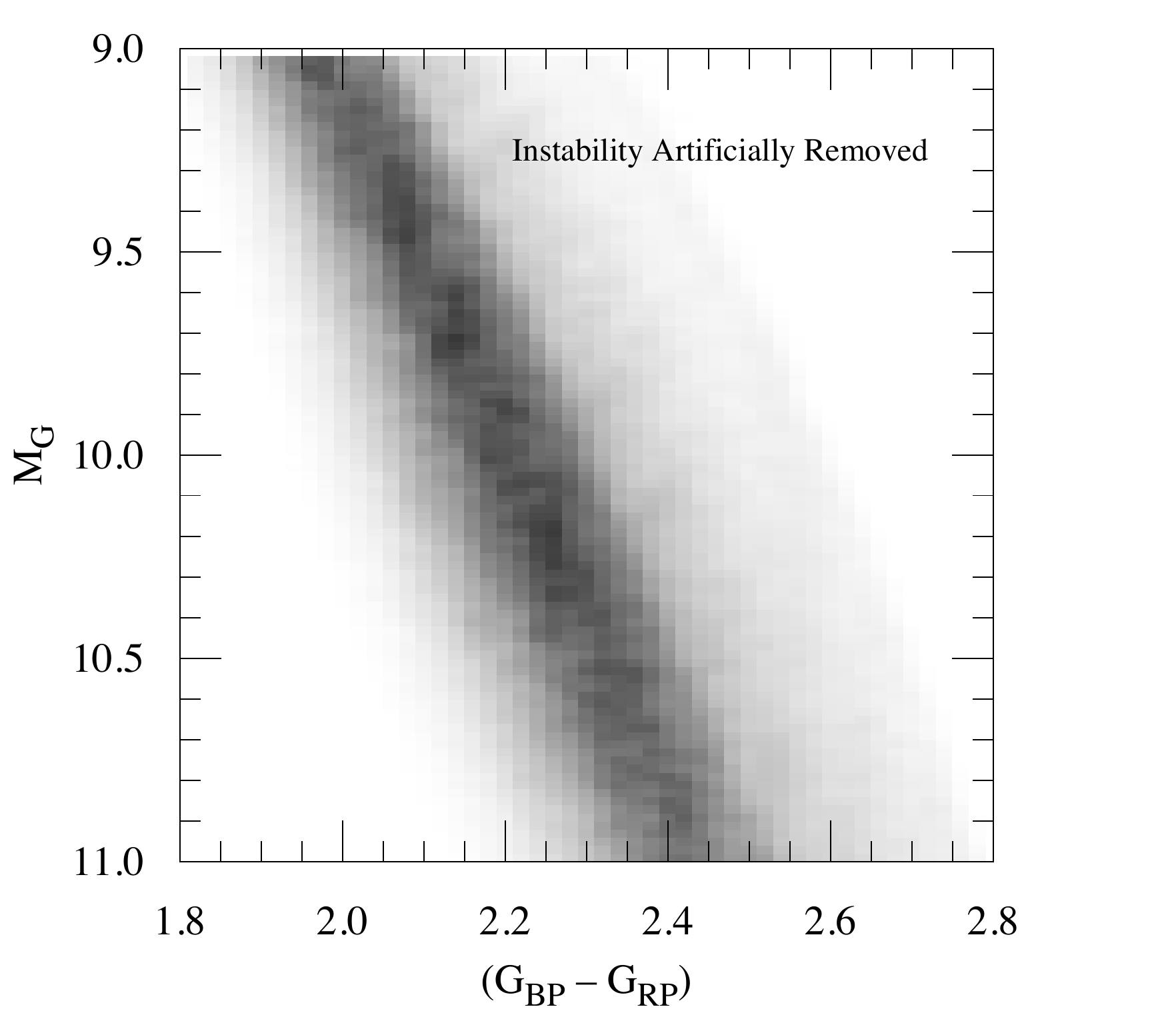}
    \caption{View of the region surrounding the Gaia M-dwarf gap in a synthetic color-magnitude diagram. Here, the \isotope[3]{He} instability has been artificially suppressed to test its impact on the synthetic CMD. Densities of stars are shown as a 2D histogram using a grey-scale.}
    \label{fig:cmd_smooth}
\end{figure}

\begin{figure*}
    \centering
    \includegraphics[width=1.75\columnwidth]{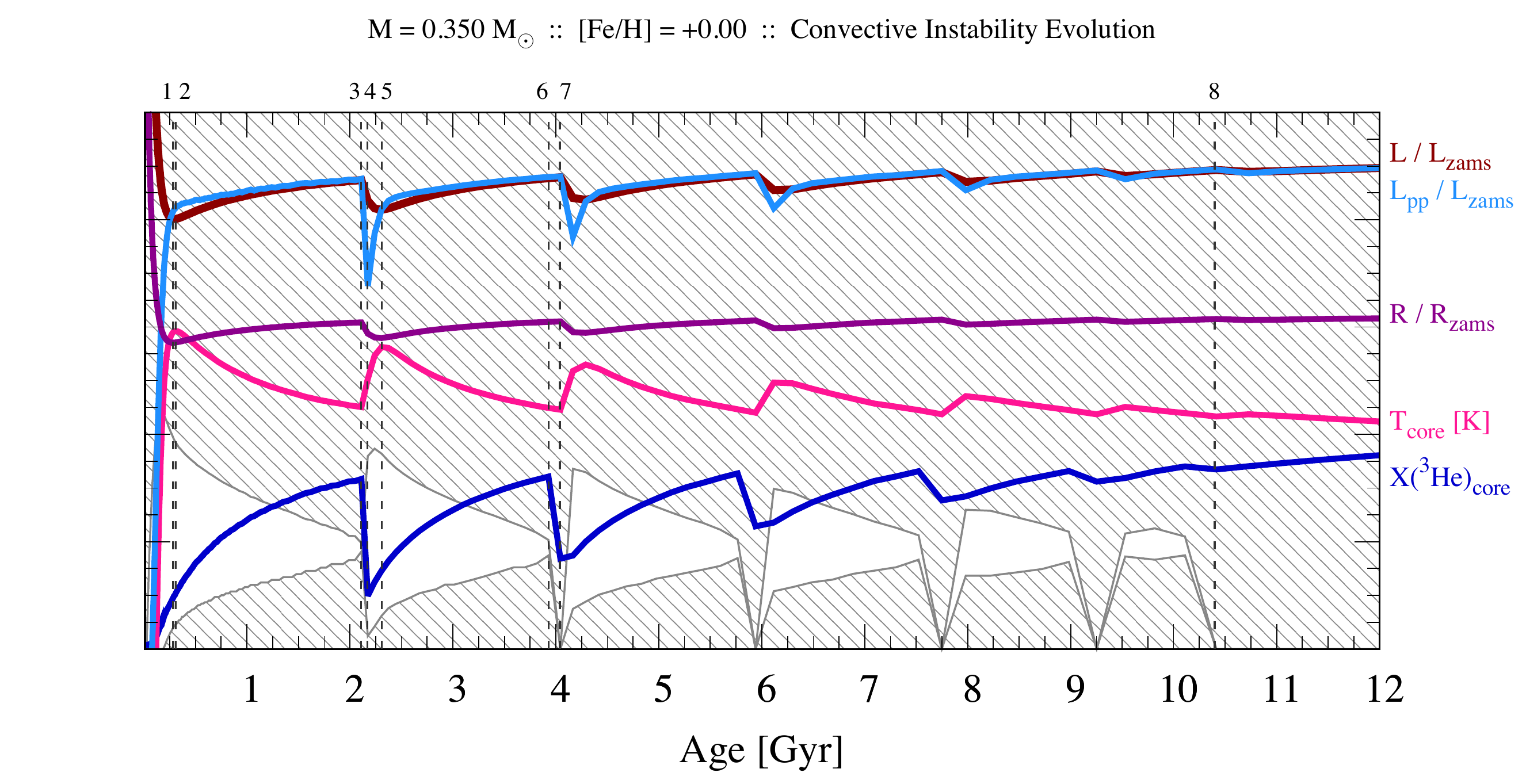}
    \caption{Kippenhan-Iben diagram for a $0.350~M_{\sun}$ model star with a solar metallicity that undergoes pulsations due to the \isotope[3]{He} fusion instability. The background shows the extent of the convective envelope and convective core throughout the star as a function of fractional stellar mass (hatched regions). Five other variables are plotted on an arbitrary $y$-axis meant to highlight how the parameters vary in time. These variables are: bolometric luminosity normalized to the zero-age main-sequence value (dark red line), luminosity from the p-p chain normalized to the zero-age main-sequence value (light blue line), stellar radius normalized to the zero-age main-sequence value (purple line), the central \isotope[3]{He} abundance (dark blue line), and the central temperature, which ranges from 6.3 -- 8.0 MK (pink line).}
    \label{fig:unstable}
\end{figure*}

Evidence for this claim comes from interior conditions in the synthetic stars. In and around the gap is the only location along the synthetic single star main sequence where there is a mixture of partially and fully convective model stars.  
As one progresses from brighter to fainter magnitude bins, there is an increasing fraction of fully convective stars. Notably, the peak of the fully convective star distribution moves toward redder colors with increasing $M_G$. Furthermore, on the red side of the gap, stars are predominantly partially convective. This is as expected if stars below the gap are fully convective and stars in the gap are in the process of transitioning. By $M_G = 10.30$, where the gap cannot be discerned in the synthetic CMD, nearly all of the stars are fully convective.

While indicative of the gap being caused by the transition to complete convection, the mixture of interior states is not a sufficient criterion to establish the \isotope[3]{He} instability as the gap's origin. 

To definitively establish the \isotope[3]{He} instability as the origin of the gap, we attempt to produce a CMD without a gap. This approach was adopted by \citet[]{Baraffe2018}. They showed that, by suppressing the \isotope[3]{He} reaction, they were able to suppress the \isotope[3]{He} instability and remove discontinuities in stellar evolution isochrones. We created a synthetic Gaia CMD from a set of stellar evolution isochrones that were smoothed to remove luminosity, radius, and effective temperature variations associated with the \isotope[3]{He} instability. The isochrones were the same in every respect to the set of isochrones used to generate the synthetic population in Figures~\ref{fig:gaia_cmd} and Figure~\ref{fig:gap_slopes}, except for the smoothing. It's also important to note that the synthetic magnitudes are calculated after stellar properties are derived from model isochrones. No smoothing is performed on the synthetic photometry. Figure~\ref{fig:cmd_smooth} shows the same region of the synthetic CMD presented in Figure~\ref{fig:gap_slopes}, where the gap should be prominent. Without variations due to the \isotope[3]{He} instability, there is no M-dwarf gap. This strongly suggests that the gap is, in fact, an observable consequence of the \isotope[3]{He} instability.

\subsubsection{van Saders Instability} \label{sec:mechanism}
Appearance of the synthetic gap is due to a structural instability that results from non-equilibrium \isotope[3]{He} fusion in stellar cores near the fully convective boundary, coined the ``convective kissing instability'' \citep{vanSaders2012}. The physics underlying the instability have been described previously \citep{vanSaders2012, Baraffe2018}. However, we provide an assessment using our models to confirm that the instability we are observing for model stars in the gap is, in fact, the same \isotope[3]{He} instability.\footnote{We cannot claim complete independence of results from \citet[]{vanSaders2012} owing to a common stellar evolution code lineage.}

The entire instability process can be understood with the Kippenhahn-Iben diagram in Figure~\ref{fig:unstable}, which summarizes the instability physics characterized previously \citep{vanSaders2012, Baraffe2018}. In the figure, the background is a standard Kippenhahn diagram. The y-axis represents the normalized stellar mass from 0 (core) to 1 (surface) with hatched areas indicating the presence of convection. Overplotted is the evolution of five different stellar properties: total stellar luminosity (dark red line), luminosity from the p-p fusion reaction (light blue line), total stellar radius (purple line), core temperature (pink line), and the core \isotope[3]{He} mass fraction (dark blue line). 
Note that the y-axis scale is different for each property plotted, although we applied a normalization and an arbitrary offset to keep values between 0 and 1. Table~\ref{tab:instability_evolution} lists values for each stellar model property at 8 separate points marked in Figure~\ref{fig:unstable}.

\begin{deluxetable*}{c c c c c c c c c l}
    %\tablewidth{2\pagewidth}
    \tablecolumns{10}
    \tablecaption{Properties of the 0.350 $M_\sun$ Model Star in Figure 5 \label{tab:instability_evolution}}
    \tablehead{\colhead{Point} & \colhead{Age} & \colhead{$M_{\rm conv.\ core}$} & \colhead{$M_{\rm conv.\ env.}$} & \colhead{$L_{\star}$} & \colhead{$L_{\rm pp}$} & \colhead{$R_{\star}$} & \colhead{$X( ^3\textrm{He})_{\rm core}$} & \colhead{$T_{\rm core}$} & \colhead{Note} \\
    \colhead{\#} & \colhead{[Gyr]} & \colhead{[$M_{\sun}$]} & \colhead{[$M_{\sun}$]} & \colhead{[$L_{\rm zams}$]} & \colhead{[$L_{\rm zams}$]} & \colhead{[$R_{\rm zams}$]} & \colhead{[$\times 10^{3}$]} & \colhead{[MK]} & \colhead{}}
    \startdata
    1 & 0.284 & 0.014 & 0.212 & 1.000 & 1.014 & 1.000 & 1.18 & 7.89 & Minimum radius ZAMS \\
    2 & 0.309 & 0.016 & 0.215 & 1.000 & 1.021 & 1.000 & 1.30 & 7.90 & 1st $T_{\rm core}$ peak \\
    3 & 2.11  & 0.064 & 0.281 & 1.072 & 1.075 & 1.039 & 3.94 & 7.69 & Pre conv.\ zone merger 1 \\
    4 & 2.17  & 0.009 & 0.224 & 1.034 & 0.876 & 1.017 & 1.23 & 7.77 & Post conv.\ zone merger 1 \\
    5 & 2.31  & 0.021 & 0.223 & 1.017 & 1.018 & 1.009 & 1.82 & 7.85 & 2nd $T_{\rm core}$ peak \\ 
    6 & 3.93  & 0.060 & 0.279 & 1.076 & 1.078 & 1.040 & 4.00 & 7.69 & Pre conv.\ zone merger 2 \\
    7 & 4.04  & 0.000 & 0.350 & 1.078 & 1.079 & 1.041 & 2.09 & 7.68 & Post conv.\ zone merger 2 \\
    8 & 10.4  & 0.000 & 0.350 & 1.091 & 1.092 & 1.045 & 4.16 & 7.67 & Fully convective
    \enddata
\end{deluxetable*}

Production of deuterium begins in earnest through p-p fusion when core temperatures exceed a few million Kelvin \citep{Salpeter1952}.\footnote{Neglecting initial deuterium fusion on the pre-main-sequence.} Deuterium's lifetime against proton capture is exceedingly short compared to stellar evolutionary timescales, meaning that the deuterium abundance is always in equilibrium and that the newly produced deuterium is immediately converted to \isotope[3]{He} \citep{Salpeter1952}. Just as p-p reactions are starting, plasma in the stellar core is becoming increasingly ionized, reducing the radiative opacity and leading to the development of a radiative core \citep[e.g.,][]{Iben1965}. This is seen at ages younger than 0.25 Gyr in Figure~\ref{fig:unstable}, where the convective envelope recedes from the core (gray hatched area). Reactions that destroy \isotope[3]{He} are negligible until core temperatures reach approximately 8 MK \citep{Parker1964}. Because the core is radiative, production of \isotope[3]{He} happens in a stably stratified environment isolated from other regions of the star where the \isotope[3]{He} abundance is lower \citep[i.e., the outer envelope;][]{vanSaders2012}. This allows the core \isotope[3]{He} abundance to increase independent of the rest of the star, as shown by the solid, dark blue line in Figure~\ref{fig:unstable}.

Disequilibrium \isotope[3]{He} fusion occurs as the model settles onto the main sequence \citep{Ezer1967,vanSaders2012}. During this time, models predict that the core is still contracting as the star proceeds toward establishing hydrostatic and thermal equilibrium. Nominally, thermal equilibrium will be established once the energy production rate in the stellar core equals the star's luminosity. However, reaction sequences that convert \isotope[3]{He} to \isotope[4]{He} contribute about half of the total energy for the full p-p chain \citep{Parker1964}. This means that, while \isotope[3]{He} fusion is in disequilibrium, the core will continue to contract \citep{Iben1965}, increasing the core temperature and the rate of \isotope[3]{He} production through the first two steps in the p-p chain \citep{Ezer1967}. 

Once the core temperature exceeds 7.7 MK, destruction of \isotope[3]{He} begins primarily through the $\isotope[3]{He} + \isotope[3]{He} \rightarrow \isotope[4]{He} + 2p$ reaction channel \citep{Parker1964}. Destruction of \isotope[3]{He} occurs at a much slower rate than \isotope[3]{He} production while the \isotope[3]{He} abundance is below its equilibrium value, thereby allowing the \isotope[3]{He} abundance to continuing growing \citep{Baraffe2018}. Meanwhile, ignition of the full PP-I chain leads to an overproduction of energy in the core, causing a convective core to develop \citep[as shown in Figure~\ref{fig:unstable};][]{Ezer1967, vanSaders2012, Baraffe2018}. The excess energy is partially released as work to expand the core and the star as a whole \citep[similar to what occurs on the post-main-sequence;][]{Iben1965,Ezer1967}. This results in a cooling of the core temperature, which causes the opacity to begin rising and the surface convective envelope to deepen. Meanwhile, the convective core continues to grow as energy production by the p-p chain exceeds the total stellar luminosity \citep{Baraffe2018}. The two convection zones continue to increase in size until either: (1) the p-p chain reaches equilibrium and stabilizes the star, or (2) they merge and the star becomes fully convective.

For stars above the Gaia M-dwarf gap, the p-p chain will begin producing enough energy to stabilize the star before the convection zones merge. Over time, the p-p chain comes into equilibrium and the convective core steadily subsides, resulting in a star with a radiative core and a convective outer envelope --- a partially convective star. However, for stars between 0.30~$M_\sun$ and 0.40~$M_\sun$, the two convection zones merge before the star stabilizes \citep[see Figure~\ref{fig:unstable};][]{vanSaders2012}.

When the two convection zones merge, plasma from the envelope and core mix \citep{vanSaders2012, Baraffe2018}. The core \isotope[3]{He} abundance decreases because the outer stellar layers had a significantly lower \isotope[3]{He} abundance than the \isotope[3]{He}-enriched core. The sudden reduction in \isotope[3]{He} causes the rate of \isotope[3]{He} + \isotope[3]{He} fusion to subside, causing a noticeable drop in the p-p chain luminosity \citep[as observed in Figure~\ref{fig:unstable}; also see][]{vanSaders2012}. With fusion now producing significantly less energy, the star contracts, releasing gravitational potential energy in the process. Thus, the total luminosity shows a less dramatic decrease than the p-p chain luminosity in Figure~\ref{fig:unstable} \citep{Baraffe2018}. At the same time, the surface and core convection zones shrink, separating once again. With the star contracting, the core temperature begins to increase in an effort to reestablish fusion reactions that consume \isotope[3]{He}. The process, as described above, repeats itself 4 -- 6 times until mergers of the convective core and envelope do not drastically alter the core \isotope[3]{He} abundance \citep{vanSaders2012, Baraffe2018}.

\subsubsection{Metallicity Dependence} \label{sec:metal_depend}
The Gaia CMD shown in Figure~\ref{fig:gaia_cmd} suggests that there is a metallicity dependence affecting the formation of the M dwarf gap. Evidence for this dependence includes the positive correlation between the gap's absolute magnitude, $M_G$, and \bprp\ color across the CMD, and the fact that the gap is more apparent at bluer colors than at redder colors. In this section, we investigate the metallicity dependence of the \isotope[3]{He} instability.

The instability occurs at roughly the same bolometric luminosity, $\log_{10}(L/L_\sun) = -1.825 \pm 0.025$, for metallicities between $-0.7$~dex -- $+0.3$~dex. At higher metallicities, the instability occurs at a significantly higher bolometric luminosity, $\log_{10}(L/L_\sun) = -1.74 \pm 0.02$. Despite model stars in the gap having broadly similar luminosities, the instability's location in absolute Gaia G-band magnitude, $M_G$, exhibits a positive, non-linear correlation with metallicity. At the same time, models undergoing the instability have higher effective temperatures at lower metallicity than models with super-solar metallicities. These relationships produce the gap's positive slope in the synthetic CMD, as hinted at by \citet{vanSaders2012}.  

\begin{deluxetable}{c c c c c}
    \tablewidth{\columnwidth}
    \tablecolumns{5}
    \tablecaption{Properties of Stars in the \isotope[3]{He} Instability Region \label{tab:unstable_masses}}
    \tablehead{
    \colhead{[Fe/H]} & \colhead{$M_{\rm low}$} & \colhead{$M_{\rm high}$} & \colhead{$\Delta M$} & \colhead{$\log_{10}(L/L_\sun)_{\rm high}$} \\
    \colhead{[dex]} & \colhead{[$M_{\sun}$]} & \colhead{[$M_{\sun}$]} & \colhead{[$M_{\sun}$]} & \colhead{---} }
    \startdata
    $-0.70$  &   0.3125   &   0.3375   &   0.025 & $-1.77$ \\
    $-0.60$  &   0.3125   &   0.3375   &   0.025 & $-1.79$ \\
    $-0.50$  &   0.3125   &   0.3375   &   0.025 & $-1.80$ \\
    $-0.40$  &   0.3175   &   0.3425   &   0.025 & $-1.80$ \\
    $-0.30$  &   0.3175   &   0.3425   &   0.025 & $-1.82$ \\
    $-0.20$  &   0.3175   &   0.3475   &   0.030 & $-1.82$ \\
    $-0.10$  &   0.3175   &   0.3475   &   0.030 & $-1.84$ \\ 
    $+0.00$  &   0.3225   &   0.3525   &   0.030 & $-1.85$ \\
    $+0.10$  &   0.3275   &   0.3625   &   0.040 & $-1.85$ \\
    $+0.20$  &   0.3275   &   0.3675   &   0.040 & $-1.84$ \\
    $+0.30$  &   0.3325   &   0.3725   &   0.040 & $-1.85$ \\
    $+0.40$  &   0.3525   &   0.3975   &   0.040 & $-1.74$ \\
    $+0.50$  &   0.3575   &   0.4025   &   0.040 & $-1.74$
    \enddata
    \tablecomments{Uncertainties: $\sigma_{M_{\rm low}} = \sigma_{M_{\rm high}} = \pm 0.0025 M_\sun$, $\sigma_{\Delta M} = \pm 0.0050 M_\sun$, $\sigma_{\log_{10}(L/L_\sun)} = \pm 0.01$.}
\end{deluxetable}

Models predict a positive correlation between the mass of stars at the fully convective boundary and metallicity, as shown in Table~\ref{tab:unstable_masses}. 
The mass range over which the instability occurs is narrow, $\Delta M = 0.025$ -- $0.040$~$M_\sun$.  In general, this supports observations that the gap occurs over a very narrow range in $M_G$ and $M_{K_s}$ \citep{Jao2018}. In addition, models predict that the range of masses affected by the \isotope[3]{He} instability increases with metallicity. This point appears counter-intuitive given that the gap is more apparent along the blue edge of the Gaia CMD more sub-solar metallicity stars are located \citep{Jao2018, Jao2020}. 

The stellar mass-metallicity correlation at the fully convective boundary is due to the dependence of radiative opacity on metallicity and the radiative opacity's effect on the outer convection zone depth \citep{Chabrier1997}. The primary sources of opacity are bound-free (b-f) and free-free (f-f) absorption with electron scattering contributing about 1\% of the total opacity \citep[e.g.,][]{Clayton1968}. Electron scattering is, therefore, not expected to strongly influence the interior structure of M dwarf stars. 
A conceptual understanding is gained by assuming that opacity sources obey Kramers' opacity law \citep{Kramers1923}.\footnote{ Hydrogen and helium are completely ionized in a non-degenerate, ideal gas.} The mean free-free (f-f) contributions follow the relationship 
\begin{equation}
    \bar{\kappa}_{ff} \propto 10^{22}\cdot(1 - Z)(1 + X)\rho T^{-7/2},
    \label{eq:opac_ff}
\end{equation}
where $X$ and $Z$ are the hydrogen and metal mass fractions, $\rho$ is the local mass density, and $T$ is the local gas temperature \citep{Clayton1968}. Bound-free (b-f) contributions are proportional to $Z$,
\begin{equation}
    \bar{\kappa}_{bf} \propto 10^{25}\cdot Z(1 + X)t^{-1}\rho T^{-7/2}.
    \label{eq:opac_bf}
\end{equation}
where $t$ is the guillotine factor, which has a typical value of $t \sim 10$ \citep{Clayton1968}. Because the total mean radiative opacity (bound-free + free-free) depends on the metal mass fraction, $Z$, there is a non-linear relationship between metallicity, [$Fe$/H] $=\log_{10}(Z/X) - \log_{10}(Z/X)_\sun$, and the mean radiative opacity $\bar{\kappa}$. 

This matters because stars in hydrostatic equilibrium must maintain a given pressure gradient and, similarly, a given temperature gradient. Increasing the radiative opacity makes radiative energy transport less efficient. Outgoing photons are more readily scattered or absorbed, steepening the radiative gradient and favoring the onset of convection in stellar interiors. As a result, models predict that stars of a given mass will have a deeper convection zone at higher metallicity. Because the relationship between metallicity and radiative opacity is non-linear, the effect of metallicity in determining the mass at the fully convective boundary is expected to be greatest at super-solar metallicities. This is in qualitative agreement with results presented in Table~\ref{tab:unstable_masses}.

Looking closer, the correlation between [Fe/H] and the maximum mass in the instability region arises from the effect of opacity on the thermal structure of M dwarf atmospheres. A higher metallicity results in a greater opacity --- particularly molecular opacities --- leading to a flatter temperature gradient \citep[e.g.,][]{Allard1995, Chabrier1997, Chabrier2000}. Temperatures deep in the stellar atmosphere are predicted to increase for a fixed effective temperature, as a result. For a given effective temperature and $\log_{10}(g)$, model atmospheres predict an approximately 1\% increase in temperature at a given optical depth (e.g., $\tau_{\rm ross} = 10$) for every 0.1~dex increase in [Fe/H]. 

Crucially, our stellar structure models predict core temperatures change by approximately 0.2\% per 0.1~dex in metallicity. This latter effect has little to do with the increase in $Z$ and is almost entirely related to an increase in helium abundance, $Y$, that accompanies an increase in [Fe/H] in our models. Since the core temperature is proportional to the mean molecular weight, $T_c \propto \mu$, increasing $Y$ increases the core temperature.\footnote{An established relationship in stellar structure theory derived in most textbooks on the subject, e.g., \citet{Chandrasekhar1939,Clayton1968}.} Therefore, the core temperature is relatively insensitive to changes in $Y$ compared to changes near the base of the stellar atmosphere caused by an increase in metallicity. This means that a temperature increase near the base of the atmosphere leads to a flatter temperature gradient throughout the star \citep{Chabrier1997,Chabrier2000}.

A flatter temperature gradient favors the onset of convection because radiation cannot efficiently transport energy throughout the star \citep[i.e., a steepening of the radiative temperature gradient; see, e.g.,][]{Cox1968}. Increases in atmospheric opacity due to increasing [Fe/H] lead to deeper and more massive outer convection zones in stars, pushing the fully convective boundary to higher masses at higher metallicity \citep{Chabrier1997}. Again, because the opacity is a non-linear function of [Fe/H], this effect will be more pronounced at super-solar metallicities compared to sub-solar metallicities. 

As mentioned above, an increase in the metal mass fraction $Z$ is accompanied by an increase in helium abundance $Y$ in our models. To be clear, this is an assumption in our models --- one that can potentially be tested (see Section~\ref{sec:helium}). Effects associated with an increasing helium abundance, particularly the effects on radiative opacity, compete against the effects caused by an increasing metallicity. Increasing the helium abundance leads to an increase in mean molecular weight and, as a consequence, a higher plasma temperature at a given pressure. A higher helium abundance also leads to a decrease in the hydrogen abundance for a given metal mass fraction $Z$. As noted in Equations~\ref{eq:opac_ff} and \ref{eq:opac_bf}, increasing the plasma temperature and decreasing $X$ both contribute to an overall lower radiative opacity. 

Overall, the relative impact of $Y$ on increasing the core temperature and decreasing the radiative opacity is small compared to the effect metallicity has on increasing the temperature deep in the stellar atmosphere. The cumulative result is that the masses of stars near the fully convective boundary exhibit a positive, non-linear correlation with metallicity.

\begin{figure*}
    \centering
    \includegraphics[width=1.6\columnwidth]{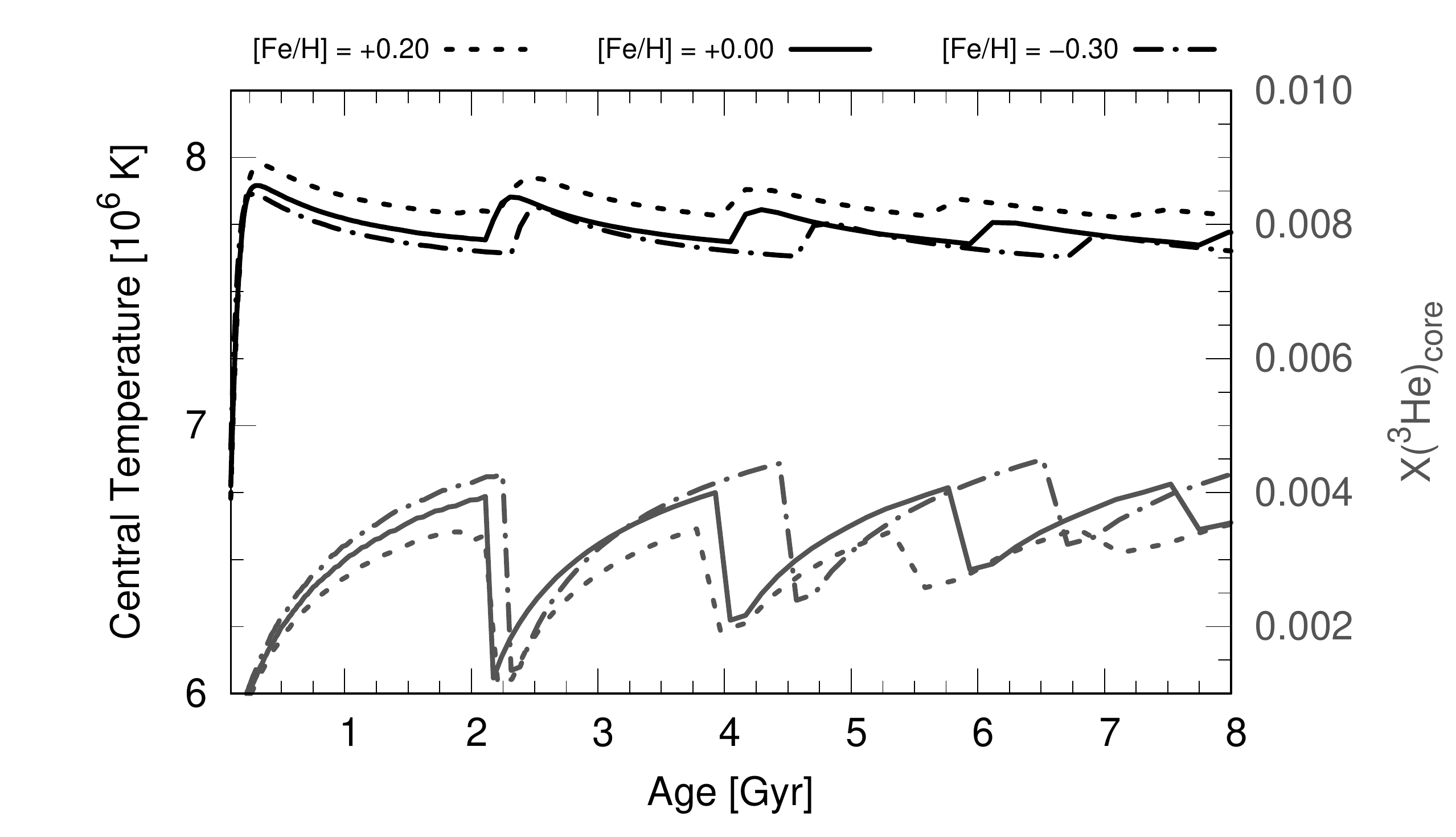}
    \caption{Predicted central temperature (black lines) and core $X(\isotope[3]{He})$ evolution (gray lines) for three model stars in the synthetic M dwarf gap. The models have different masses and metallicities: ([Fe/H], $M/M_\sun$) = ($-0.3$, $0.34$), ($+0.0$, $0.35$), ($+0.20$, $0.36$) and are represented by a dot-dashed, continuous, and dashed line, respectively.}
    \label{fig:gap_central_T}
\end{figure*}

These results do not immediately reveal why the M dwarf gap is more prominent at bluer colors. Looking at a theoretical Hertzsprung-Russell diagram ($\log(L/L_\sun)$ vs.\ $T_{\rm eff}$), one finds that the gap is already more prominent at warmer temperatures compared to cooler temperatures. This suggests that interior physics are more responsible than atmospheric physics. Comparing models of stars predicted to undergo the largest pulsations at a given metallicity, we find that sub-solar metallicity models have a longer pulsation period by approximately 200 Myr compared to a solar metallicity model. In addition, sub-solar metallicity models show pulsation amplitudes that are about 1\% larger in bolometric luminosity than solar metallicity models. The opposite appears to be true for model effective temperatures: increasing the metallicity leads to larger pulsation amplitudes. However, changes in effective temperature are on the sub-1\% level (roughly 10 K), compared to changes in bolometric luminosity, which changes by 5\% or more.

Sub-solar metallicity models show larger variations in bolometric luminosity than super-solar metallicity models because of differences in their core temperature. This is illustrated in Figure~\ref{fig:gap_central_T}. There are two effects that conspire to produce the result. First is that stellar core temperatures are directly proportional to the plasma's mean molecular weight. A higher metallicity and, more importantly, a higher helium abundance, produces a hotter core temperature.\footnote{Recall that the model helium abundance is tied directly to metallicity.} Effectively, to exert a given pressure --- in this case, the one required to maintain hydrostatic equilibrium --- the plasma must be hotter when the average particle mass is larger. Near the fully convective boundary, the mean molecular weight changes by approximately 0.6\% per 0.1 dex in metallicity, which corresponds a comparable change in core temperature of about 0.4\% per 0.1 dex in metallicity. Second is that models predict the mass of stars in the gap is directly proportional to metallicity as a result of higher radiative opacities (see above). The core temperature of stars at the fully convective boundary changes by about 2\% per 0.01 $M_\sun$.

Each of the noted physical effects implies that super-solar metallicity models in the gap have higher core temperatures than sub-solar metallictiy models. With a higher core temperature, the abundance of \isotope[3]{He} required to initiate and sustain the p-p chain decreases. Assuming a sub-solar and super-solar metallicity star in the gap have the same initial abundance of \isotope[3]{He}, it will take a lower-metallicity model longer to build up a sufficient amount of \isotope[3]{He} to establish the complete p-p chain fusion reaction sequence, leading to longer pulsation periods and larger pulsation amplitudes. The net result is that lower-metallicity models are predicted to undergo larger and longer pulsations, producing a more noticeable gap on the blue side of the Gaia main sequence.

\section{Discussion} \label{sec:discussion}

\subsection{Comparison to Previous Work} \label{sec:previous}
Our modeling confirms that the Gaia M dwarf gap is the result of successive mergers between the convective core and envelope of low-mass main sequence stars. These merging events reduce the core \isotope[3]{He} abundance, which reduces the p-p chain luminosity, thereby decreasing the model star's luminosity. This mechanism, described in Section~\ref{sec:mechanism}, is the same mechanism originally described by \citet[]{vanSaders2012} and further elucidated by \citet[]{Baraffe2018}. However, the mechanism favored by our modeling disagrees with the mechanism proposed by \citet[]{MacDonald2018}. 

The two proposed mechanisms depend on the core abundance of \isotope[3]{He}, but in different ways. For one mechanism, the core \isotope[3]{He} abundance decreases during successive core-envelope merging events \citep[][this work]{vanSaders2012, Baraffe2018}. This mechanism predicts that the core \isotope[3]{He} abundance increases while the convective core and envelope are separated due to non-equilibrium \isotope[3]{He} fusion in the core (production exceeds destruction). Meanwhile, the abundance of \isotope[3]{He} in the convective envelope increases only marginally, while the two regions are separated. In contrast, the second mechanism posits that the core \isotope[3]{He} abundance increases until it reaches a quasi-equilibrium value, while the envelope \isotope[3]{He} abundance is permitted to increase above this equilibrium value \citep{MacDonald2018}. A single merging event is proposed to occur, which leads to an increased core \isotope[3]{He} abundance. 

There could be multiple reasons why there is some disagreement about the gap formation mechanism. Likely candidates include: (1) differences in numerical schemes used to solve the stellar structure equations, (2) assumptions in solving the nuclear reaction network, and (3) treatment of convective mixing
\citep[instantaneous vs diffusive;][respectively]{vanSaders2012, MacDonald2018}. Convective mixing treatments, highlighted by \citet[]{MacDonald2018}, should only play a minor role and are unlikely to be the culprit \citep{Chabrier1997, Baraffe2018}.
Nevertheless, it is not clear how each of these differences affect the stellar evolution calculations in the vicinity of the \isotope[3]{He} instability --- no direct comparison has been performed --- but the end result appears to be different predictions about whether \isotope[3]{He} reaches a quasi-equilibrium in the stellar core \citep{MacDonald2018, Baraffe2018}. 
Our modeling suggests that \isotope[3]{He} does not have adequate time to reach a quasi-equilibrium in the stellar core. This is supported by estimates of \isotope[3]{He} lifetimes against fusion with another \isotope[3]{He} nucleus ($\sim10^9$ yr), which is the only relevant reaction at temperatures between 3 -- 7 MK \citep[][]{Parker1964, Clayton1968}. Models from \citet[]{MacDonald2018} reach quasi-equilibrium in 300 -- 500 Myr (based on their Figure 3), which is difficult to reconcile with estimates of  the \isotope[3]{He} destruction lifetime, as noted by \citet[]{Baraffe2018}. 

There is a general consensus on the mass regime where the gap is formed at solar metallicity: between 0.31~$M_\sun$ and 0.37~$M_\sun$. However, there are differences among the modeling groups about the exact mass range where the gap is formed. Our models predict the gap will occur for stars with masses between 0.315 -- 0.355~$M_\sun$. Models from the Yale Rotating Evolution Code \citep[YREC;][]{vanSaders2012} predict that the gap will be produced by stars with masses between 0.322 and 0.365~$M_\sun$, while \citet[]{Baraffe2015} models predict a more narrow mass range of 0.34 -- 0.36 $M_\sun$ \citep{Baraffe2018}. Despite modeling differences and disagreements about the proposed instability mechanism, \citet[]{MacDonald2018} predict a similar mass range of 0.310 -- 0.345~$M_\sun$ for the masses of stars in the gap. 

Similar results are found if we consider the luminosity of unstable stars, instead of mass. Three out of the four model sets predict a solar metallicity model star with the largest amplitude \isotope[3]{He} instability has a luminosity $\log_{10}(L/L_\sun) \sim -1.85$ \citep[][this work]{MacDonald2018, Baraffe2018}, while models from \citet[]{vanSaders2012} predict the instability at a higher luminosity around $\log_{10}(L/L_\sun) \sim -1.75$. 

Given that we found no difference between model sets using a \citetalias{GS98} and \citetalias{GAS07} solar composition, we rule out the possibility that the higher solar $Z$ adopted by \cite[]{vanSaders2012} is responsible for the luminosity difference. By running a series of test models, we were also able to rule out the optical depth fitting point for surface boundary conditions and the adopted convective mixing length parameter ($\alpha_{\rm mlt}$) as explanations for the luminosity difference. This leaves the equation of state (OPAL vs. FreeEOS) and the radiative opacity tables (OP vs OPAL) as plausible explanations (\citet{vanSaders2012} vs this work, respectively). As noted previously, the exact set of adopted macro- and micro-physics can alter the predicted gap location \citep{vanSaders2012, Baraffe2018}. Further tests are needed to isolate the impact of different physical ingredients on the synthetic gap properties. 

\begin{figure*}[t]
    \centering
    \includegraphics[width=1.9\columnwidth]{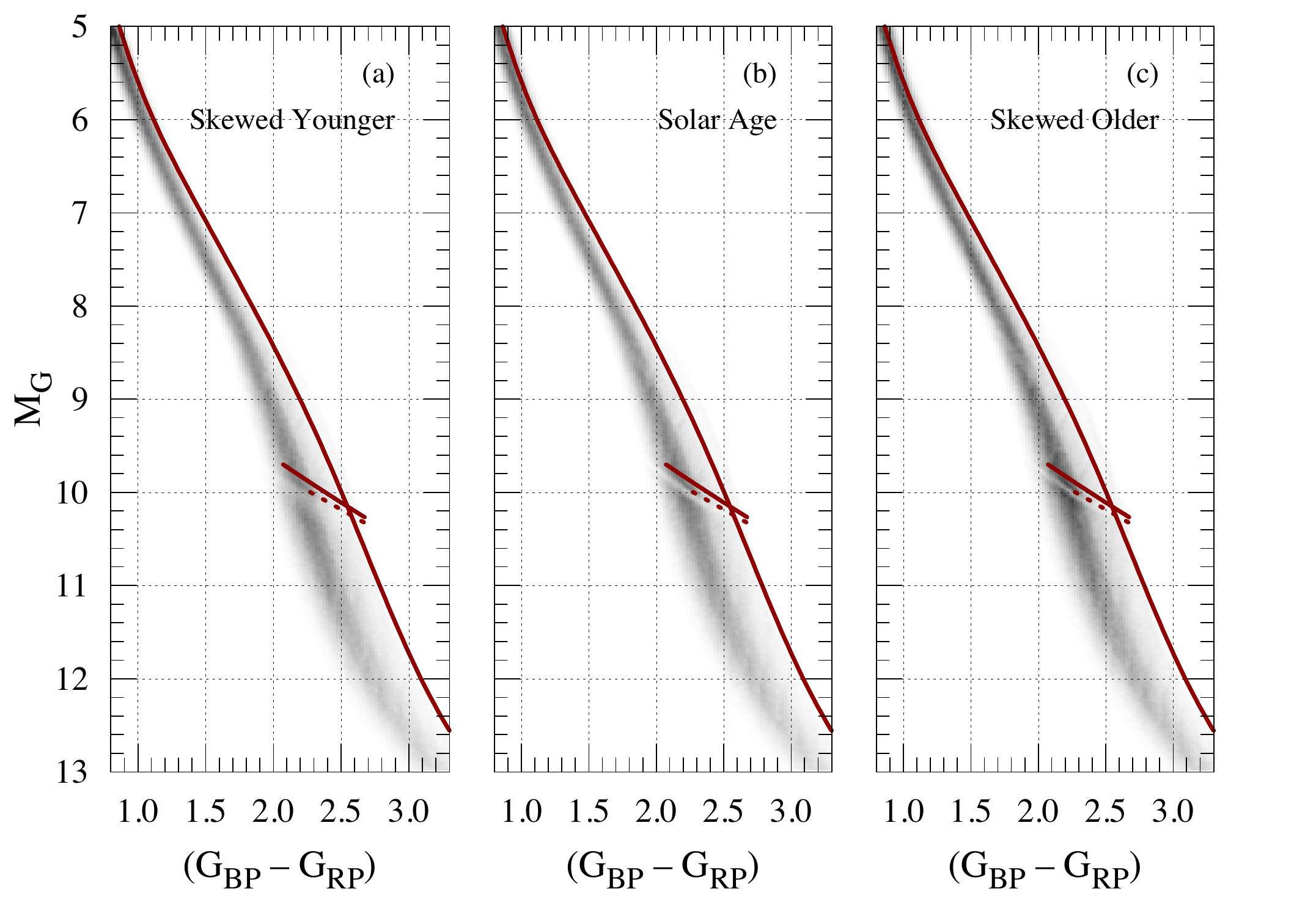}
    \caption{Synthetic Gaia CMDs created assuming that the age distribution is: (a) skewed toward stars younger than 5 Gyr, (b) normally distributed around 5 Gyr, and skewed toward stars older than 5 Gyr. Red lines show the locations of the observed Gaia main sequence spine, the observed bright-edge of the M dwarf gap \citep{Jao2020}, and the observed $M_G$ of greatest decrement \citep{Jao2018}.}
    \label{fig:cmd_age_dists}
\end{figure*}

\subsection{Age Distribution Sensitivity} \label{sec:ages}
Our synthetic CMD in Figure~\ref{fig:gaia_cmd} was created assuming that M dwarf stars in the solar neighborhood have ages uniformly distributed between 0.5 Gyr and 10 Gyr. To test the impact of this assumption on our results, we created three additional synthetic stellar populations with different age distributions between 0.5 Gyr and 10 Gyr:
\begin{enumerate}
    \item a normal distribution with $\mu = 5$~Gyr and $\sigma = 1.2$~Gyr,
    \item a distribution skewed toward younger ages that obeys a gamma distribution $\Gamma\left(k = 1.5, \theta = 1.2\right)$, with shape $k$ and scale $\theta$ given in Gyr,
    \item a distribution skewed toward older ages that obeys a gamma distribution $\Gamma\left(k = 8.5, \theta = 1.2\right)$, with $k$ and $\theta$ given in Gyr.
\end{enumerate}
Figure~\ref{fig:cmd_age_dists} shows synthetic Gaia CMDs created using each of these age distributions. The CMDs look similar outside of the Gaia M dwarf gap. They each have a prominent main sequence spine near the peak of the metallicity distribution and they all have similar main sequence widths. This is not entirely unexpected. M dwarf stars do not evolve significantly after they reach the main sequence (after about 300 Myr), meaning the precise age distribution does not contribute to significant differences in the overall synthetic CMD \citep{Laughlin1997}. The metallicity distribution plays a larger role in governing the overall appearance of the CMD.

There are differences near the M dwarf gap. With an age distribution skewed younger (Figure~\ref{fig:cmd_age_dists}(a)), the M dwarf gap has softer edges and an apparently larger vertical extent in $M_G$. The gap also appears to not be as under dense compared to the surrounding CMD as in the other cases. In contrast, age distributions that are normally distributed around 5 Gyr (Figure~\ref{fig:cmd_age_dists}(b)) or skewed toward stars older than 5 Gyr (Figure~\ref{fig:cmd_age_dists}(c)) produce results very similar to a uniform distribution (Figure~\ref{fig:gaia_cmd}). The M dwarf gaps are prominent with sharp, well-defined edges, especially on the bluer edge of the main sequence. There is also a trend that the M dwarf gaps are less vertically extended in $M_G$ as the age distribution moves toward older ages. However, in each case, the M dwarf gap has a location and slope similar to the original CMD presented in Figure~\ref{fig:gaia_cmd}. 

The M dwarf gap's appearance with age is associated with the \isotope[3]{He} instability's age dependence. Recall from Section~\ref{sec:results} that the highest amplitude pulsations --- the ones expected to produce features noticeable in a CMD --- are predicted to have periods of 1 -- 2 Gyr and occur 4 -- 6 times in a star's lifetime. Pulsations can occur on timescales of several hundred million years, but the pulsation amplitudes are relatively small and do not produce significant variability in a CMD. Therefore, an age distribution skewed toward younger ages is less likely to capture stars undergoing high amplitude pulsations. This means that there will be more stars present in the gap, yielding a less dramatic under density compared to other scenarios. But, because low amplitude pulsations occur on shorter timescales and typically cease to occur after 2 Gyr, there will be more stars undergoing pulsations at younger ages, producing a wider gap.  Each of these effects is captured in the synthetic CMD for the young M dwarf distribution (Figure~\ref{fig:cmd_age_dists}(a)), which has an M dwarf gap that is less well-defined, but more vertically extended than in the other three cases. 

\begin{figure}
    \centering
    \includegraphics[width=0.75\columnwidth]{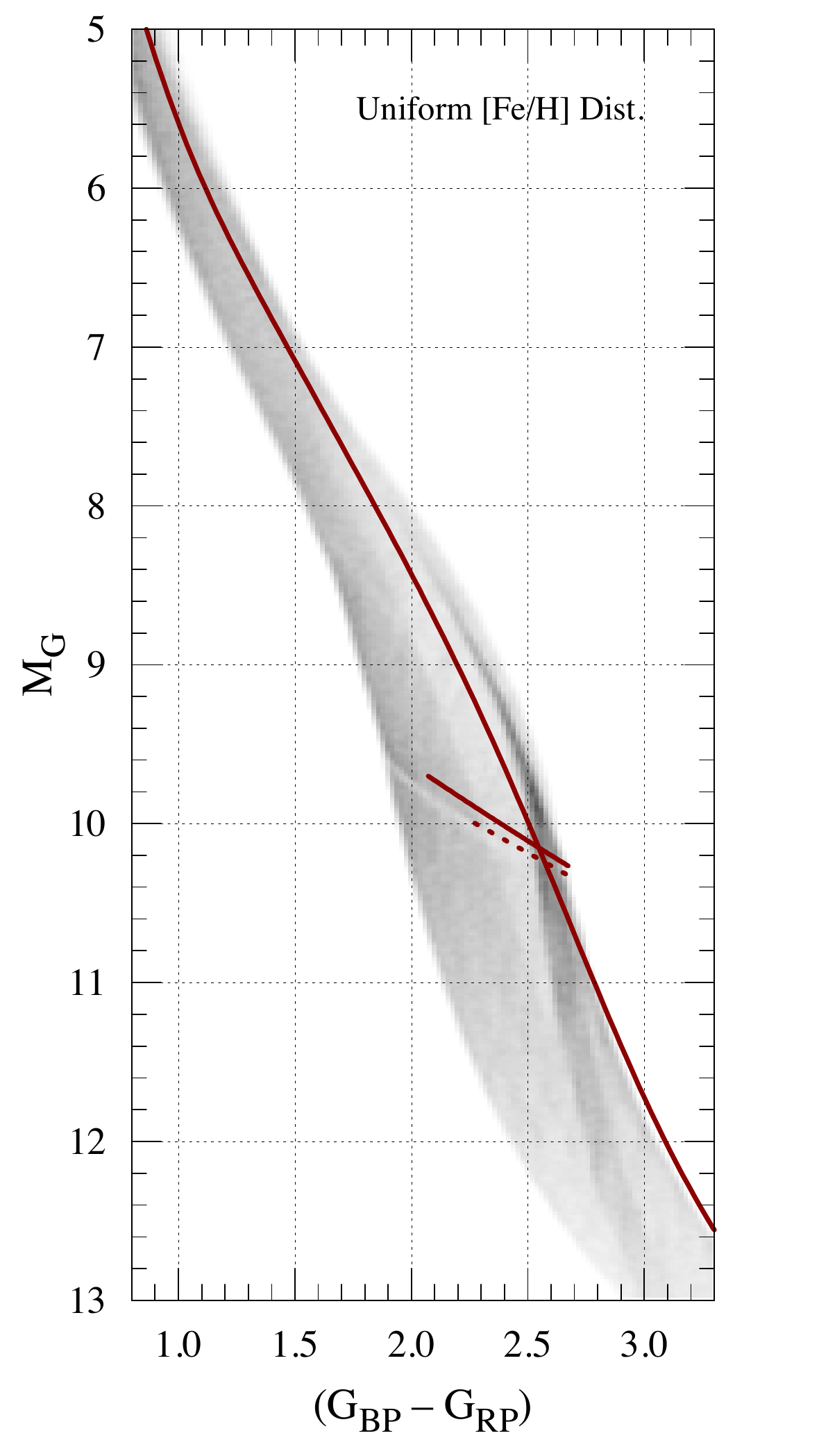}
    \caption{Synthetic Gaia CMD created using an uniform metallicity distribution with $-0.7 \le$~[Fe/H]~$\le +0.50$. Red lines show the locations of the observed Gaia main sequence spine, the observed bright-edge of the M dwarf gap \citep{Jao2020}, and the observed $M_G$ of greatest decrement \citep{Jao2018}.}
    \label{fig:cmd_uni_feh}
\end{figure}

A more detailed analysis of the gap as a function of the age distribution or including a more realistic star formation history in our population synthesis model may reveal subtle differences. These differences could be used to constrain the Milky Way's star formation history. In the meantime, the precise age distribution does not significantly affect our results. 

\subsection{Metallicity Distribution Sensitivity} \label{sec:feh}
Our simple population synthesis assumed that metallicities were normally distributed following results from the Geneva-Copenhagen Survey \citep{Nordstrom2004}. We tested the impact of this choice by generating a synthetic CMD with a metallicity distribution uniform between $-0.7 \le$~[Fe/H]~$\le +0.5$. The resulting CMD is shown in Figure~\ref{fig:cmd_uni_feh}.

Adopting a uniform metallicity distribution removes the appearance of a prominent main sequence spine. The main sequence spine was attributed to the metallicity distribution peak in Sections~\ref{sec:results} and \ref{sec:ages}, and here we see evidence in support of that idea. There is an apparent over-density among the reddest M dwarf stars in the CMD extending from $M_G \sim 8$ through $M_G \sim 13$.  We do not currently have an explanation for this feature. However, it is important to note that we do not see such an over-density in the observed Gaia CMD \citep{Jao2020}. Whether the over-density in the synthetic CMD is the result of errors in model atmospheres or a valid prediction from metal-rich stellar structure models is not clear. Critically, errors with predictions for high metallicity stars would have a minimal impact on the results in Section~\ref{sec:results} because these stars exist in the tail ($> 3\sigma$ from the mean) of the adopted metallicity distribution. They also do not contribute to the appearance of the M dwarf gap, which appears most prominent at bluer colors among solar and sub-solar metallicity model stars.

Regardless of the over-dense feature among the high metallicity stars, the morphology of the Gaia M dwarf gap remains consistent with the synthetic gap described in Section~\ref{sec:results}. Therefore, we conclude that the choice of underlying metallicity distribution does not affect our results. 

\subsection{Wider Implications}
%% Can the gap be used for "galactic archaeology" with M dwarfs?
Sections~\ref{sec:mechanism} and \ref{sec:metal_depend} revealed that the synthetic gap's morphology is sensitive to the core conditions within M dwarf stars near the fully convective boundary. This implies that the Gaia M dwarf gap may be used to help the constrain physical ingredients in stellar structure models and, potentially, to constrain the local star formation history.

\subsubsection{M Dwarf Interior Physics} \label{sec:m_dwarf_physics}
The sensitivity of the \isotope[3]{He} instability to conditions in the stellar core suggests that the gap is a powerful feature to benchmark model physics. In particular, the gap may prove useful as a diagnostic for validating explanations for the M dwarf radius problem, where stellar models predict M dwarf radii that are 5 -- 10\% smaller than radii measured using a variety of techniques \citep[e.g.,][]{Ribas2006, Torres2010, Feiden2012a, Boyajian2012, Spada2013, Mann2015, Morrell2019}. While this problem has been attributed to rapid rotation, there is evidence that the problem may be independent of rotation rate \citep{Kesseli2018, Morrell2019}.

Multiple explanations have been proposed with the leading hypothesis being that magnetic fields or activity (i.e., starspots) are responsible \citep[e.g.,][]{LopezMorales2007}. Models attempting to include these physics have demonstrated that model radii can be brought into agreement with observations \citep{Mullan2001, Chabrier2007, Feiden2013, Feiden2014, Somers2015}. Also finding success are models that include an ad-hoc adjustment to the atmospheric thermal structure for M dwarfs \citep{Chen2014}, chosen to bring model properties into agreement with observed M dwarf properties. Critically, these model sets typical predict small changes to the M dwarf mass-luminosity relation \citep[e.g.,][]{Feiden2013}, meaning the core conditions with stellar models adopting non-standard physical ingredients may alter the location of the Gaia M dwarf gap in a synthetic CMD.

Previous tests have shown that changing model effective temperatures and radii also brings model photometry into agreement with observed CMDs \citep[e.g.,][]{Chen2014, Morrell2019}. However, the gap provides an independent check on the model physics by revealing the core conditions within these model stars. We have performed such an analysis and find that, while models are able to simultaneously correct fundamental properties and synthetic photometry \citep{Chen2014, Morrell2019}, the predict a that the gap is about 0.3 -- 0.5 magnitudes fainter than the observed gap. Models can either get surface properties correct or the core properties correct, but not both simultaneously. Our analysis will be presented in a forthcoming paper. 

\subsubsection{Helium Evolution} \label{sec:helium}
Another avenue to explore is the relationship between stellar metal mass fraction, $Z$, and helium mass fraction, $Y$. Stellar evolution models typically prescribe a linear relationship between these two quantities
\begin{equation}
    Y = Y_{\rm prim} + \frac{\Delta Y}{\Delta Z}\cdot \left( Z - Z_{\rm prim}\right),
\end{equation}
where $Y_{\rm prim}$ is the primordial helium mass fraction following big bang nucleosynthesis, $Z_{\rm prim}$ is the metal mass fraction (typically taken to be zero), and $\Delta Y / \Delta Z$ is a characteristic slope describing how $Y$ evolves with $Z$ throughout cosmic time. The latter is generally fixed to a value determined by the calibration of a standard solar model. Attempts to measure $\Delta Y / \Delta Z$ provide conflicting opinions about the value of this quantity, typically with $ 1 \lesssim \Delta Y / \Delta Z \lesssim 6$ \citep[e.g.,][and references therein]{Fernandes1996, Casagrande2007, Verma2019}. 

The synthetic Gaia M dwarf gap occurs at a nearly constant bolometric luminosity and is a sensitive function of the core conditions in stellar models (as seen, e.g., in Section~\ref{sec:metal_depend}). Therefore, variations in the helium abundance with metallicity will likely alter the morphology of the gap in a synthetic CMD, particularly the slope of the gap and its vertical extend in $M_G$. This sensitivity makes the gap a useful feature for understanding the evolution of helium with metallicity, i.e., $\Delta Y / \Delta Z$. There is some concern that the continual mixing occurring within fully convective stars would bias the results toward estimating higher present-day helium abundances than at the time of the star's formation. However, the effect is small ($\Delta Y \le 0.007$) compared to variations predicted by differences in $\Delta Y / \Delta Z$.

It remains to be seen whether other physical uncertainties would prohibit useful restrictions on the helium evolution. However, an extensive study exploring different values for $\Delta Y / \Delta Z$ may provide an independent constraint on the range of permitted values for this quantity.  %although it remains to be seen whether other uncertainties would dominate any effect produced by a varying helium abundance.

\subsubsection{Local Star Formation History} \label{sec:star_formation}
Sections~\ref{sec:ages} and \ref{sec:feh} demonstrated that the average magnitude $\langle M_G\rangle$ and slope of the M dwarf gap was largely unaffected by the choice of age and metallicity distribution. However, alternative distributions did have observable consequences. For example, the age distribution appeared to affect the contrast of the gap with the rest of the main sequence and the gap's height in absolute magnitude, $M_G$. By contrast, the adopted metallicity distribution appeared to affect the presence and, potentially, the location of the main sequence spine. 

Therefore, we speculate that the properties of the Gaia M dwarf gap may provide insight into the age and metallicity distribution of the local M dwarf population. In particular, the observed gap exhibits a triangular shape at colors bluer than the main sequence spine \citet[]{Jao2020}, a feature that our synthetic populations struggle to reproduce. A combination of at least two separate age distributions signifying major star formation episodes, with at least one component skewed toward younger ages, may alleviate this difficulty. 

\section{Conclusions}
We used simple population synthesis along with a dense grid of low-mass stellar evolution models to test whether non-equilibrium \isotope[3]{He} fusion \citep{Ezer1967, vanSaders2012} was responsible for the formation of the Gaia M dwarf gap \citep{Jao2018, Jao2020}. Our synthetic CMD exhibited a gap near the fully convective boundary that was quantitatively similar to the observed gap. The synthetic gap was located at bluer colors and a brighter magnitudes than the observed gap, in line with current M dwarf modeling trends \citep[e.g.,][]{Chen2014}. However, the synthetic gap had a slope nearly identical to the observed gap's slope across the CMD. 

We confirmed that a structural instability due to non-equilibrium \isotope[3]{He} fusion \citep{vanSaders2012} is responsible for the synthetic gap by artificially removing variations in stellar properties due to the instability. Without variations in stellar properties due to the \isotope[3]{He} instability, there is no M dwarf gap in the synthetic CMD. Combining this result with the agreement between slopes for the observed and synthetic M dwarf gap, this suggests that the metallicity dependence of the \isotope[3]{He} instability in stellar models is reasonably accurate. 

We also found that the synthetic gap appears more prominent at bluer colors, consistent with observations \citep{Jao2020}. We conclude that this is due to the dependence of \isotope[3]{He} instability's amplitude and duration on predicted stellar core temperatures \citep[previously suggested by][]{vanSaders2012, Baraffe2018}. In particular, we propose that the dependence of core temperatures on stellar mass and helium abundance (via the mean molecular weight) are the critical factors. Because the mass of stars at the fully convective boundary is directly proportional to metallicity, bluer stars in the gap have a lower mass and cooler core temperatures, resulting in larger amplitude and longer duration pulsations due to the \isotope[3]{He} instability. 

The Gaia M dwarf gap is a novel feature in the CMD and provides some of the first insight into the core conditions of M dwarf stars. This creates new opportunities to study M dwarf physics and to utilize M dwarfs for probing galactic evolution. In particular, the gap will provide a critical test of different hypotheses about why model M-dwarf stars have smaller than expected radii. The gap allows for the first simultaneous tests of M dwarf atmospheric physics (colors and magnitudes) and interior physics (non-equilibrium \isotope[3]{He} fusion). While further work is needed to understand the gap's sensitivity to assumptions in stellar evolution modeling, it should prove to be a critical benchmark for stellar models.

\acknowledgements
K.S. thanks A.~Ash and J.~Hamilton for their guidance, and is grateful for support from the Ronald E. McNair Post-Baccalaureate Achievement Program and the National Science Foundation S-Stem Award No.\ 1458731. G.A.F.\ thanks B.\ Edvardsson, T.\ Vogel, and A.~W.\ Mann for fruitful discussions on multiple elements of the paper. G.A.F.\ is also grateful for support from a Presidential Semester Incentive Award from the University of North Georgia and the CoSMS Institute at the University of North Carolina, Chapel Hill for their hospitality. W.-C.J.  is supported by the National Science Foundation under grant AST-1715551.

\software{FreeEOS \citep{FreeEOS},
          Gnuplot \citep{Gnuplot},
          NumPy \citep{NumPy}, 
          SciPy \citep{SciPy},
          Matplotlib \citep{matplotlib},
          lmfit (\url{https://github.com/lmfit/lmfit-py})
          }

%%%%%%%%%%%%%%%%%%%%%%%%%%%%%%%%%%%%%%%%%%%%%%%%%%



%\bibliographystyle{mnras}

%\bibliography{}{}
\bibliographystyle{aasjournal}
\bibliography{ms.bib}

\end{document}